\documentclass[11pt]{article}

\usepackage{jheppub} 
\usepackage{slashed}
\usepackage[T1]{fontenc} 
\usepackage{amsmath,amsfonts,amsthm,bm}
\usepackage{braket}
\usepackage{stackengine}
\usepackage{mathtools}
\usepackage{physics}
\usepackage{xcolor}
\usepackage{hhline} 
\usepackage{multirow}

\DeclareMathOperator{\arctanh}{arctanh}

\renewcommand{\vec}{{}}
\newcommand{\vvec}{\mathbf}

\newcommand{\AD}{A^{\dagger}}
\newcommand{\AT}{\widetilde{A}}
\newcommand{\ATD}{\widetilde{A}^{\dagger}}
\newcommand{\BT}{\widetilde{B}}
\newcommand{\BTD}{\widetilde{B}^{\dagger}}
\newcommand{\biloc}[3]{#1(\vec{k}_{#2}, \vec{k}_{#3})}
\newcommand{\bilocS}[5]{#1(#4\vec{k}_{#2}, #5\vec{k}_{#3})}

\newcommand{\DD}[2]{\delta^{d}(\vec{k}_{#1}-\vec{k}_{#2})}
\newcommand{\Intk}{\int \mathrm{d}^d k \:}
\newcommand{\Hc}[1]{\hat{\mathcal{H}}_{2}(\vec{k}_{#1})}
\newcommand{\Gc}[1]{\hat{\mathcal{G}}_{2}(\vec{k}_{#1})}
\newcommand{\Oc}[1]{\mathcal{O}_{#1}}
\newcommand{\COMM}[2]{[#1,#2]}
\newcommand{\ACOMM}[2]{\{#1,#2\}}
\newcommand{\sgn}[2]{s_{#2}(\mathcal{O}_{#1})}

\allowdisplaybreaks

\title{Dynamical Symmetry and the Thermofield State at Large $N$}

\author[a]{Antal Jevicki,}
\author[a]{Xianlong Liu,}
\author[b]{Junggi Yoon}
\author[a]{and Junjie Zheng}

\affiliation[a]{Department of Physics, Brown University, \\ 182 Hope Street, Providence, RI 02912, USA}
\affiliation[b]{Asia Pacific Center for Theoretical Physics, \\ Postech, Pohang 37673, Korea}

\emailAdd{antal\_jevicki@brown.edu}
\emailAdd{xianlong\_liu@brown.edu}
\emailAdd{junggi.yoon@apctp.org}
\emailAdd{junjie\_zheng@brown.edu}

\abstract{We discus Thermofield Double QFT at real time, in the large $N$ limit. First, we establish a (dynamical) symmetry which we argue holds in general on the real time portion of the Schwinger-Keldysh contour. At large $N$ this symmetry is seen to generate a one parameter degeneracy of stationary collective solutions. The construction is explicitly worked out on the example of $O(N)$ vector QFT. As a nontrivial application we describe construction of the corresponding (large $N$) Thermofield Double State in real time collective formalism.}

\keywords{Thermofield Double, Large $N$, Collective Field Theory, $O(N)$ Vector Model}

\allowdisplaybreaks

\makeatletter
\gdef\@fpheader{\vspace{10pt}}
\makeatother

\begin{document}

\maketitle
\flushbottom

\section{Introduction}

Thermal physics forms the basis for major applications of QFT in condensed matter, quantum gravity, and AdS/CFT. A theoretical formulation is given through quantization on the Schwinger--Keldysh (SK) contour \cite{Schwinger:1960qe, Keldysh:1964ud} and the corresponding perturbative \cite{Herzog:2002pc, Horava:2020she, Haehl:2016pec, Nair:2015rsa} and non-perturbative \cite{Israel:1976ur, Maldacena:2001kr} features. One particular picture is based on the thermofield double (TFD), a scheme which identifies the (two) different Hamiltonians corresponding to propagation on the Real and Imaginary parts of the SK contour, respectively. A central role is then played by the thermofield double state, which in this framework is analogous to the ground state in standard quantization. However, in contrast to canonical ground states, the thermofield double state is much less developed. Attempts at constructing and approximating the TFD state are numerous, and a partial list of references is \cite{Cottrell:2018ash, Martyn:2018wli, Wu:2018nrn, Berenstein:2019yfv}. 
It is the goal of the present work to provide a contribution to this subject, through the role (and construction) of a dynamical symmetry which we argue operates in general (for any theory) on the real-time portion of the SK contour. We demonstrate this in perturbation theory. The second ingredient is the use of large $N$. It is expected in general that in this limit the dynamics are simplified, and that an explicit construction of the TFD state might be possible. We indeed give such a construction in the collective (Hamiltonian) framework using the $O(N)$ theory as an example. We build on the initial work of \cite{Jevicki:2015sla} where the collective Hamiltonian representation of the real time SK contour was used to study emergent \cite{Jevicki:2015pza} space--time. A number of interesting recent works concern the $O(N)$ model \cite{Banerjee:2019iwd, Engelsoy:2021fbk} and its holographic role. The dynamical symmetry and the large $N$ collective TFD state wavefunction(al) described in the present work clearly hold more general relevance. Studies of TFD as a ground state through an explicitly coupled system \cite{Maldacena:2018lmt} are numerous \cite{Garcia-Garcia:2019poj, Plugge:2020wgc, Alet:2020ehp}. The formulation that we present potentially offers another line of comparison.

This work is organized as follows. In Section \ref{sec:G-symmetry}, we perturbatively establish a dynamical symmetry of the interacting thermofield theory. The field transformations under this symmetry are then discussed and are seen to be simplified at large $N$. In Section \ref{sec:collective-field-theory}, we present the full large $N$ construction in the framework of collective field theory. It presents a simple, explicit form of the dynamical symmetry in question. In Section \ref{sec:TFD-state}, we provide the construction of the TFD state. Conclusions and relevant applications are discussed in Section \ref{sec:conclusion}.

\section{Symmetry of Thermofield Double QFT}
\label{sec:G-symmetry}

We will perform our study of thermofield dynamics in the $O(N)$ QFT framework. In the first subsection we set up the notations and briefly summarize the free theory case. Generalizing to interacting QFT, we then argue for the existence of an exact dynamical thermofield double symmetry. We give a demonstration of this, and a construction of the generator of thermofield symmetry in perturbation theory. The associated symmetry transformations of fields are also given. The role of this symmetry will be visible in the solution of large $N$ collective field theory that follows in the subsequent section. In general, the existence of symmetry implies a degeneracy in the stationary field solution. This will be seen through the collective field theory solution of the $O(N)$ QFT.

\subsection{Thermofield Dynamics in Free Theory}

In the TFD formalism, a thermofield vacuum state $| 0(\beta) \rangle$ is defined so that the thermal average of an arbitrary operator $\mathcal{O}$ can be reproduced as an expectation value in the thermal vacuum state $| 0(\beta) \rangle$
\begin{equation}
    \langle \mathcal{O} \rangle_{\beta} \,\equiv\, \langle 0(\beta) | \mathcal{O} | 0(\beta) \rangle \,=\, \frac{1}{Z(\beta)} \operatorname{Tr}(e^{-\beta H} \mathcal{O}) ,
\end{equation}
where $Z(\beta)\,\equiv\,\Tr(e^{-\beta H})$. To accomplish this, one should purify the mixed state by doubling the Hilbert space. Let $d$ denote the spatial dimension; the real-time thermofield Hamiltonian for the free $O(N)$ vector model is
\begin{equation}
    \hat{H}_{2} \,=\, 
    \int \left[ \frac{1}{2} \pi^2 + \frac{1}{2}(\nabla \varphi)^2 + \frac{1}{2} m^2 \varphi^2\right] \mathrm{d}^{d} x - 
    \int \left[ \frac{1}{2} \widetilde{\pi}^2 + \frac{1}{2}(\nabla \widetilde{\varphi})^2 + \frac{1}{2} m^2 \widetilde{\varphi}^2\right] \mathrm{d}^{d} x.
\end{equation}
Here $\varphi^2 \equiv \sum_{i=1}^{N} \varphi^i \varphi^i$, etc. The non-tilde (tilde) fields are the original (doubled) degrees of freedom, and we will always put a `hat' 
on the thermofield Hamiltonian and the subsequent symmetry operator~$\hat{G}$. Since we focus on the Hamiltonian formalism, for compactness we will use $x\equiv \mathbf{x}$ to denote the spatial vector, and $k \equiv \mathbf{k}$ to denote the spatial momentum. The mode expansions of the fields are as usual
\begin{equation}
    \varphi^j (\vec{x}) \,=\, \int \frac{\mathrm{d}^{d} k}{(2\pi)^{d/2}} \frac{1}{\sqrt{2 \omega(\vec{k})}} \Big(a^{j}(\vec{k})e^{i \vec{k}\cdot\vec{x}} + a^{j\dagger}(\vec{k})e^{-i\vec{k}\cdot\vec{x}}\Big),
\end{equation}
\begin{equation}
    \pi^j (\vec{x}) \,=\, - i \int \frac{\mathrm{d}^{d} k}{(2\pi)^{d/2}} \sqrt{\frac{\omega(\vec{k})}{2}} \Big( a^{j}(\vec{k})e^{i \vec{k}\cdot\vec{x}} + a^{j\dagger}(\vec{k})e^{-i\vec{k}\cdot\vec{x}} \Big),
\end{equation}
and similarly for the doubled fields $\widetilde{\varphi}$ and $\widetilde{\pi}$. 
We have the canonical commutation relations 
\begin{align}
    & [a^{i}(\vec{k}), a^{j\dagger}(\vec{k}^{\prime})] \,=\, \delta^{i  j}\delta^d(\vec{k} - \vec{k}^{\prime}), \label{eq:commutator-a-a-dagger}\\
    & [\widetilde{a}^{i}(\vec{k}), \widetilde{a}^{j\dagger}(\vec{k}^{\prime})] \,=\, \delta^{i j}\delta^{d}(\vec{k} - \vec{k}^{\prime}). \label{eq:commutator-at-at-dagger}
\end{align}
Using the mode expansion, we can write the real-time thermofield Hamiltonian as
\begin{equation}\label{eq:H2}
    \hat{H}_{2} = \Intk \omega(\vec{k}) \hat{\mathcal{H}}_2(\vec{k}) = \int \mathrm{d}^{d}k \: \omega(\vec{k}) \Big(a^{\dagger}(\vec{k}) \cdot a(\vec{k}) - \widetilde{a}^{\dagger}(\vec{k}) \cdot \widetilde{a}(\vec{k})\Big).
\end{equation}

The thermo vacuum state for the $O(N)$ model then reads
\begin{equation}
    | 0(\beta) \rangle \,\equiv\, \operatorname{exp}(- i \hat{G}_2) |0, \tilde{0} \rangle
    \,=\, \prod_{\vec{k}} \exp[\theta_{\beta}(\vec{k})\big(a^{\dagger}(\vec{k})\cdot \widetilde{a}^{\dagger}(\vec{k}) - a(\vec{k}) \cdot \widetilde{a}(\vec{k})\big)]|0, \tilde{0}\rangle.
\end{equation}
Here, the $\hat{G}_2$ operator takes the form
\begin{equation}\label{eq:G2}
    \hat{G}_{2} \,=\, \Intk i \theta_{\beta}(\vec{k}) \hat{\mathcal{G}}_2(\vec{k}) = \int \mathrm{d}^{d}k \: i \theta_{\beta}(\vec{k}) \Big(a^{\dagger}(\vec{k})\cdot \widetilde{a}^{\dagger}(\vec{k}) - a(\vec{k}) \cdot \widetilde{a}(\vec{k})\Big),
\end{equation}
with
\begin{equation}
    \theta_{\beta}(\vec{k}) \,\equiv\, \arctanh(e^{-\beta\omega(\vec{k})/2}).
\end{equation}
The $\hat{G}_2$ operator induces a Bogoliubov transformation
\begin{equation}
    a_{\beta}^{j} (\vec{k}) \equiv e^{i \hat{G}_2} a^{j} (\vec{k}) e^{-i \hat{G}_2} = 
    a^{j}(\vec{k}) \cosh\theta_{\beta}(\vec{k}) - \widetilde{a}^{j \dagger}(\vec{k}) \sinh\theta_{\beta}(\vec{k}),
\end{equation}
and similarly for $a_{\beta}^{j\dagger}(\vec{k})$, $\widetilde{a}_{\beta}^{j}(\vec{k})$ and $\widetilde{a}_{\beta}^{j\dagger}(\vec{k})$. The new annihilation operators annihilate the thermo vacuum state:
\begin{equation}
    a_{\beta}^{j} (\vec{k})|0(\beta)\rangle = \widetilde{a}_{\beta}^{j}(\vec{k}) |0(\beta)\rangle = 0.
\end{equation}

The thermofield Hamiltonian $\hat{H}_2$ then also annihilates $|0(\beta)\rangle$
\begin{equation}
    \hat{H}_2 |0(\beta)\rangle = 0.
\end{equation}
We can also easily check that $\hat{G}_2$ commutes with $\hat{H}_2$:
\begin{equation}\label{eq:comm-G2-H2}
    \COMM{\hat{G}_2}{\hat{H}_2} = 0.
\end{equation}
We note  that in $\hat{G}_{2}$, one can  replace $\theta_{\beta}$ with any well-defined even function $f(\vec{k})$ of momentum to obtain a symmetry operator $\hat{G}_{2}[f]$, obeying Equation \eqref{eq:comm-G2-H2}. Consequently, in what follows $\theta_\beta(\vec{k})$ will be replaced generally by $f(\vec{k})/2$.  

\subsection{Thermofield Dynamics in Interacting Theory}

We would like to suggest that the kinematic symmetry seen in the previous subsection is generalizable. In particular, we will argue and present evidence that it can be extended to the interacting case, taking the form of a dynamical symmetry:
\begin{equation} \label{eq:G-H-commute-condition}
    \left[\hat{G}[f],\hat{H}\right] = 0,
\end{equation}
associated with the (real-time) Hamiltonian $\hat{H}=H-\widetilde{H}$. In addition, we will argue that this generator gives the TFD vacuum state through
\begin{equation}
    | 0(\beta) \rangle = \exp(-i \hat{G}[f]) |0,\tilde{0}\rangle,  \quad \text{for specific} \quad f(\vec{k}) .
\end{equation}
In what follows, we will be demonstrating the existence and explicit construction of $\hat{G}[f]$ in perturbation theory. This will be performed through solving the condition \eqref{eq:G-H-commute-condition} .

Consider then the interacting theory with a quartic interaction $c \: \hat{H}_4$ added in the thermofield Hamiltonian
\begin{equation} \label{eq:H_interacting}
    \hat{H} = \hat{H}_2 + c \: \hat{H}_4 + \cdots,   \qquad \text{with} \quad
    \hat{H}_4 = \frac{1}{4 N} \int \left[(\varphi^2)^2 - (\widetilde{\varphi}^2)^2\right] \mathrm{d}^d x,
\end{equation}
where $c$ denotes the coupling constant. Consequently, the operator $\hat{G} \equiv \hat{G}[f]$ will be \footnote{We adopt the following notation 
\begin{equation*}
    [\mathrm{d}k]_4 \equiv (2\pi)^d \delta^d(\vec{k}_1 + \vec{k}_2 - \vec{k}_3 - \vec{k}_4) \prod_{i=1}^{4} \frac{\mathrm{d}^d k_i}{(2\pi)^{d/2}} \frac{1}{\sqrt{2\omega(\vec{k}_i)}}.
\end{equation*}}
\begin{equation}\label{eq:G-hat-interactive}
    \hat{G} = \hat{G}_2 + c\: \hat{G}_4 + \cdots, \qquad \text{with} \quad
    \hat{G}_4 = \frac{i}{4N} \int [\mathrm{d}k]_4 \hat{\mathcal{G}}_4 (\vec{k}_1, \vec{k}_2; \vec{k}_3, \vec{k}_4),
\end{equation}
where $\hat{G}_2$ is the free theory generator \eqref{eq:G2}. We will be constructing $\hat{G}_4$ in momentum space.
Substituting these expressions into the symmetry condition \eqref{eq:G-H-commute-condition} and using the existing commutation relation \eqref{eq:comm-G2-H2}, we see that up to order $c$, the $\hat{G}_{4}$ operator obeys the following symmetry constraint
\begin{equation}\label{eq:constraint}
    \COMM{\hat{G}_2}{\hat{H}_4} - \COMM{\hat{H}_2}{\hat{G}_4} 
    = 0.
\end{equation} 
Our goal is to solve this equation for $\hat{G}_4$, determining $\hat{G}[f]$ to this order. 

It is advantageous to work in momentum space, and with bi-local operators specified in Appendix \ref{appendix:bi-local-algebra}. We can write $\hat{G}_{2}$ and $\hat{H}_{2}$ as
\begin{align}
 i\hat{G}_2&=-\frac{1}{2}\int \mathrm{d}^d k \: f(\vec{k})(C^\dagger(\vec{k},\vec{k})-C(\vec{k},\vec{k})),\\
 \hat{H}_2&=\int \mathrm{d}^d k \: \omega(\vec{k})(B(\vec{k},\vec{k})-\widetilde{B}(\vec{k},\vec{k})).
\end{align}
Defining two linear combinations of bi-local operators
\begin{align}
    & \biloc{S}{1}{2} = \biloc{A}{1}{2} + \bilocS{\AD}{1}{2}{-}{-} + 2 \bilocS{B}{1}{2}{-}{}, \label{eq:S-operator} \\
    & \biloc{\widetilde{S}}{1}{2} = \biloc{\AT}{1}{2} + \bilocS{\ATD}{1}{2}{-}{-} + 2 \bilocS{\BT}{1}{2}{-}{}, \label{eq:S-tilde-operator}
\end{align}
we can also write the quartic interaction in terms of bi-local operators as
\begin{align}
    \nonumber
    \hat{H}_4 & = \frac{1}{4N} \int [\mathrm{d}k]_4 \hat{\mathcal{H}}_4 (\vec{k}_1, \vec{k}_2; \vec{k}_3, \vec{k}_4) \\
    & = \frac{1}{4N} \int [\mathrm{d}k]_4 \left( 
        \biloc{S}{1}{2} \bilocS{S}{3}{4}{-}{-} - 
        \biloc{\widetilde{S}}{1}{2} \bilocS{\widetilde{S}}{3}{4}{-}{-} \right).
\end{align}
The definitions and commutation relations of bi-local operators appearing in Equations \eqref{eq:S-operator} and \eqref{eq:S-tilde-operator} are summarized in Appendix \ref{appendix:bi-local-algebra}. We see that $\hat{\mathcal{H}}_{4}$ is a quadratic form of the bi-local operators, which implies that $\hat{\mathcal{G}}_4$ will also be a quadratic form.

With these expressions, after evaluating the first commutator of Equation \eqref{eq:constraint} and integrating over the momentum variables in $\hat{G}_2$ we obtain the following equation for $\hat{G}_{4}$
\begin{small}
\begin{align}\label{eq:G2-comm-H4}
        \COMM{\hat{H}_{2}}{\hat{G}_{4}} = &
        \COMM{\hat{G}_2}{\hat{H}_4} \nonumber \\
        = & -\frac{i}{4 N} \int [\dd{k}]_4 \: f(\vec{k}_2) \:
         \ACOMM
        {\biloc{D}{1}{2} + 
        \bilocS{D^{\dagger}}{1}{2}{-}{-} + 
        \bilocS{C}{1}{2}{}{-} + 
        \bilocS{C^{\dagger}}{1}{2}{-}{}}
        {\bilocS{S}{3}{4}{-}{-}} \nonumber \\
        & + \frac{i}{4 N} \int [\dd{k}]_4 \: f(\vec{k}_1) \:
        \ACOMM
        {\bilocS{D}{1}{2}{-}{-} +
        \bilocS{D^{\dagger}}{1}{2}{}{} + 
        \bilocS{C}{1}{2}{-}{} + 
        \bilocS{C^{\dagger}}{1}{2}{}{-}}
        {\bilocS{\widetilde{S}}{3}{4}{-}{-}},
    \end{align}
  \end{small}
where $\ACOMM{\cdot}{\cdot}$ denotes the anti-commutator. We discuss the solution of this equation next.

To solve this equation, we first observe that the commutation relations of $\Hc{}$ and a bi-local operator $\biloc{\Oc{}}{1}{2}$ \eqref{eq:COMM-Hc-1} to \eqref{eq:COMM-Hc-10} can be schematically put into the form
\begin{equation}
    \COMM{\Hc{}}{\mathcal{O}(\vec{k}_1, \vec{k}_2)} = s_1(\mathcal{O}) \DD{}{1} \mathcal{O}(\vec{k}, \vec{k}_2) + s_2 (\mathcal{O}) \DD{}{2} \mathcal{O}(\vec{k}_1, \vec{k}).
\end{equation}
Here, $s_1$ and $s_2$ are signs that depend on the bi-local operator $\biloc{\Oc{}}{1}{2}$. Suppressing the momentum arguments, their values can be summarized as follows
\begin{enumerate}
    \item $\sgn{}{1} = -1, \: \sgn{}{2} = -1, \: \forall\: \mathcal{O} \in \{ A, \: \ATD, \: D \}$.
    \item $\sgn{}{1} = +1, \: \sgn{}{2} = +1, \: \forall\: \mathcal{O} \in \{ \AD, \: \AT, \: D^{\dagger} \}$.
    \item $\sgn{}{1} = +1,\: \sgn{}{2} = -1, \: \forall \: \mathcal{O} \in \{ B, \: C^\dagger \}$.
    \item $\sgn{}{1} = -1,\: \sgn{}{2} = +1, \: \forall \: \mathcal{O} \in \{ \BT, \: C \}$.
\end{enumerate}
Thus, for an arbitrary quadratic form of the bi-local operators, we have
\begin{align}
    \COMM{\Hc{}}{\biloc{\Oc{a}}{1}{2}\biloc{\Oc{b}}{3}{4}} = & +
    \sgn{a}{1} \DD{}{1} \biloc{\Oc{a}}{}{2} \biloc{\Oc{b}}{3}{4} \nonumber \\
    & +
    \sgn{a}{2} \DD{}{2} \biloc{\Oc{a}}{1}{} \biloc{\Oc{b}}{3}{4} \nonumber \\
    & +  
    \sgn{b}{1} \DD{}{3} \biloc{\Oc{a}}{1}{2} \biloc{\Oc{b}}{}{4} \nonumber \\
    & +
    \sgn{b}{2} \DD{}{4} \biloc{\Oc{a}}{2}{2} \biloc{\Oc{b}}{3}{}.
\end{align}
Let $\{\vec{k}_i\}$ be the shorthand for $\{\vec{k}_1, \vec{k}_2; \vec{k}_3, \vec{k}_4\}$; we define
\begin{equation}
    \omega_{\Oc{a}\Oc{b}}(\{\vec{k}_i\}) \equiv \sgn{a}{1}\omega(\vec{k}_1) + \sgn{a}{2}\omega(\vec{k}_2) + \sgn{b}{1}\omega(\vec{k}_3)+\sgn{b}{2}\omega(\vec{k}_4),
\end{equation}
then for any quadratic form of bi-local operators we have the identity
\begin{equation}
    \frac{\mathcal{F}(\{\vec{k}_i\})}
    {\omega_{\Oc{a}\Oc{b}}(\{\vec{k}_i\})}
    \COMM{\hat{H}_2}{\biloc{\Oc{a}}{1}{2}\biloc{\Oc{b}}{3}{4}}
    =
    \mathcal{F}(\{\vec{k}_i\})
    \biloc{\Oc{a}}{1}{2} \biloc{\Oc{b}}{3}{4}.
\end{equation}
Here, $\mathcal{F}(\{\vec{k}_i\})$ is a well-defined function of the momenta which may depend on the inverse temperature $\beta$. This identity then trivially extends to
\begin{equation}\label{eq:H2-ac-quad}
    \frac{\mathcal{F}(\{\vec{k}_i\})}
    {\omega_{\Oc{a}\Oc{b}}(\{\vec{k}_i\})}
    \COMM{\hat{H}_2}{\ACOMM{\biloc{\Oc{a}}{1}{2}}{\biloc{\Oc{b}}{3}{4}}}
    =
    \mathcal{F}(\{\vec{k}_i\})
    \ACOMM{\biloc{\Oc{a}}{1}{2}}{\biloc{\Oc{b}}{3}{4}}.
\end{equation}

With Equations \eqref{eq:G2-comm-H4} and \eqref{eq:H2-ac-quad}, we see that the solution of Equation \eqref{eq:constraint} can be written as
\begin{align}
    \hat{\mathcal{G}}_4 (\vec{k}_1, \vec{k}_2; \vec{k}_3, \vec{k}_4) =
    & - f(\vec{k}_2) 
    \sum_{\alpha,\beta} \frac{1}{\omega_{\Oc{\alpha}\Oc{\beta}}(\vec{k}_1, \vec{k}_2; \vec{k}_3, \vec{k}_4)}
    \ACOMM{\Oc{\alpha}(\vec{k}_1, \vec{k}_2)}{\Oc{\beta}(\vec{k}_3, \vec{k}_4)} \nonumber \\
    & + f(\vec{k}_1)
    \sum_{\alpha,\beta} \frac{1}{\omega_{\Oc{\alpha}\widetilde{\mathcal{O}}_{\beta}}(\vec{k}_1, \vec{k}_2; \vec{k}_3, \vec{k}_4)}
    \ACOMM{\Oc{\alpha}(-\vec{k}_1, -\vec{k}_2)}{\widetilde{\mathcal{O}}_{\beta}(\vec{k}_3, \vec{k}_4)}, \label{eq:G4-schematical}
\end{align}
where the bi-local operators are labeled as
\begin{align*}
   & \mathcal{O}_{\alpha}(\vec{k}_1, \vec{k}_2) \in  
   \{ \biloc{D}{1}{2}, \:
   \bilocS{D^{\dagger}}{1}{2}{-}{-}, \:
   \bilocS{C}{1}{2}{}{-}, \:
   \bilocS{C^{\dagger}}{1}{2}{-}{} \}, \\
   & \mathcal{O}_{\beta}(\vec{k}_3, \vec{k}_4) \in
   \{ \bilocS{A}{3}{4}{-}{-}, \:
      \bilocS{\AD}{3}{4}{}{}, \:
      2 \bilocS{B}{3}{4}{}{-}
       \}, \\
   & \widetilde{\mathcal{O}}_{\beta}(\vec{k}_3, \vec{k}_4) \in
   \{ \bilocS{\AT}{3}{4}{-}{-}, \:
      \bilocS{\ATD}{3}{4}{}{}, \:
      2 \bilocS{\BT}{3}{4}{}{-}
       \}.
\end{align*}

\subsection{Field Transformations}

In this subsection we discuss the field transformations generated by $\hat{G}$. We will explicitly evaluate their form to the leading order in $c$. We consider the transformations of the (bi-local) fields induced by the $\hat{G}$ operator \eqref{eq:G-hat-interactive} through
\begin{align}
\label{eq:field-transformation-G-operator}
\Phi_f&=e^{-i\hat{G}}\Phi e^{+i\hat{G}} \nonumber \\
 &=e^{-i\hat{G}_2}\Phi e^{i\hat{G}_2}+c \: \left. \partial_c\left(e^{-i\hat{G}}\Phi e^{i\hat{G}}\right)\right|_{c=0}+\cdots .
\end{align}
The second term on the last line can be further expressed as
\begin{equation}
     \left.\partial_c\left(e^{-i\hat{G}}\Phi e^{i\hat{G}}\right)\right|_{c=0}= -\int_0^1 \dd{x} [i\hat{G}_4(x), \Phi(1)] ,
\end{equation}
where we denote
\begin{equation}
     O(x)=e^{-ix\hat{G}_2}O e^{ix\hat{G}_2}
\end{equation}
for the arbitrary operator $O$.

To evaluate the field transformations in the large $N$ limit, which is our interest, we use the algebra of bi-local operators. From the interaction contribution to the generator $\hat{G}$ eq. \eqref{eq:field-transformation-G-operator}, we have
\begin{small}
\begin{align}
 i\hat{G}_4 = & \frac{1}{2}\int \prod_{j=1}^{4}\left(\frac{\dd[d]{k_j}}{(2\pi)^d}\frac{1}{\sqrt{2\omega_j}}\right)(2\pi)^d\delta^d \left(\sum_{j=1}^{4}k_j \right)f(k_1) \nonumber \\
& \times \left\{ 
    \left[ -\frac{1}{\omega_1+\omega_2+\omega_3+\omega_4}D(k_2,k_1)A(k_3,k_4)+\frac{1}{\omega_1+\omega_2-\omega_3-\omega_4}D^\dagger(-k_2,-k_1)A(k_3,k_4)\right.\right. \nonumber \\
 & \quad -\frac{1}{\omega_1-\omega_2+\omega_3+\omega_4}C^\dagger(-k_2,k_1)A(k_3,k_4)+\frac{1}{\omega_1-\omega_2-\omega_3-\omega_4}C(k_2,-k_1)A(k_3,k_4) \nonumber \\
 & \quad \left.\left.
    -\frac{2}{\omega_1+\omega_2-\omega_3+\omega_4}D(k_2,k_1)B(-k_3,k_4)+\frac{2}{\omega_1-\omega_2-\omega_3+\omega_4}C(k_2,-k_1)B(-k_4,k_3)
    \right]\right. \nonumber \\
 & \quad - \left[\dagger\right]\bigg{\}} + \left\{ \thicksim\right\}.
\end{align}
\end{small}
Here, $\dagger$ represents adjoint conjugation, and $\thicksim$ represents the analogous contributions in terms of the tilde operations.

Likewise, for the bi-local fields:
\begin{align}
    \Phi_{11}(k_1,k_2)&=\frac{1}{\sqrt{2\omega_1}\sqrt{2\omega_2}}\left\{A(k_1,k_2)+B(-k_1,k_2)\right\}+\{\dagger\}, \nonumber \\
    \Phi_{12}(k_1,k_2)&=\frac{i}{\sqrt{2\omega_1}\sqrt{2\omega_2}}\left\{C(k_1,k_2)+D(k_1,-k_2)\right\}+\{\dagger\} .
\end{align}
We note that even though singular terms appear in the above, these will be seen to cancel among each other, leaving a non-singular result for the resulting field transformations. This represents a rather nontrivial consistency check on the construction and form of the symmetry generator to this order.

We now give the symmetry transformation of $\Phi_{11}$, with other components being similar. We find it convenient to exhibit the form in an integral representation over $x$:
\begin{small}
\begin{align}
&-c\int_0^1 \dd{x} [i\hat{G}_4(x), \Phi_{11}(p,q;1)] \nonumber \\
= &
-\frac{c}{2}\frac{1}{\sqrt{2\omega_p}\sqrt{2\omega_q}}\int_0^1 \dd{x} \int \prod_{j=1}^4\left(\frac{\dd[d]{k_j}}{(2\pi)^d}\frac{1}{\sqrt{2\omega_j}}\right) (2\pi)^d\delta^d(\sum_{j=1}^4 k_j) f(k_1)\times \nonumber \\
&
\left\{+\frac{2}{\omega_1-\omega_2+\omega_3+\omega_4}\left(\cosh{x\frac{f_2}{2}}\sinh{x\frac{f_1}{2}}B_{\bar{2}1}+\sinh{x\frac{f_2}{2}}\cosh{x\frac{f_1}{2}}\tilde{B}^\dagger_{\bar{2}1}\right)\times\right. \nonumber \\
&
\left[\cosh{x\frac{f_3}{2}}\cosh{x\frac{f_4}{2}}\cosh{\frac{f_p}{2}}\cosh{\frac{f_q}{2}}(\delta_{4\bar{p}}B_{\bar{q}3} + \delta_{4\bar{q}}B_{\bar{p}3}) - \sinh{x\frac{f_3}{2}}\sinh{x\frac{f_4}{2}}\sinh{\frac{f_p}{2}}\sinh{\frac{f_q}{2}}(\delta_{4\bar{p}}\tilde{B}^\dagger_{\bar{q}3} + \delta_{4\bar{q}}\tilde{B}^\dagger_{\bar{p}3}) \right] \nonumber \\
&
+\frac{2}{-\omega_1+\omega_2+\omega_3+\omega_4}\left(\cosh{x\frac{f_2}{2}}\sinh{x\frac{f_1}{2}}B^\dagger_{\bar{2}1}+\sinh{x\frac{f_2}{2}}\cosh{x\frac{f_1}{2}}\tilde{B}_{\bar{2}1}\right)\times \nonumber \\
&
\left[\cosh{x\frac{f_3}{2}}\cosh{x\frac{f_4}{2}}\cosh{\frac{f_p}{2}}\cosh{\frac{f_q}{2}}(\delta_{4\bar{p}}B_{\bar{q}3} + \delta_{4\bar{q}}B_{\bar{p}3}) - \sinh{x\frac{f_3}{2}}\sinh{x\frac{f_4}{2}}\sinh{\frac{f_p}{2}}\sinh{\frac{f_q}{2}}(\delta_{4\bar{p}}\tilde{B}^\dagger_{\bar{q}3} + \delta_{4\bar{q}}\tilde{B}^\dagger_{\bar{p}3}) \right] \nonumber \\
&
+\frac{2}{\omega_1+\omega_2-\omega_3+\omega_4}\left(\cosh{x\frac{f_3}{2}}\cosh{x\frac{f_4}{2}}B_{\bar{3}4}+\sinh{x\frac{f_3}{2}}\sinh{x\frac{f_4}{2}}\tilde{B}^\dagger_{\bar{3}4}\right)\times \nonumber \\
&\hspace{3cm}
\left[+\cosh{x\frac{f_2}{2}}\sinh{x\frac{f_1}{2}}\cosh{\frac{f_p}{2}}\cosh{\frac{f_q}{2}}(\delta_{1\bar{p}}B_{\bar{q}2} + \delta_{1\bar{q}}B_{\bar{p}2}+\delta_{2\bar{p}}B_{\bar{q}1} + \delta_{2\bar{q}}B_{\bar{p}1})\right. \nonumber \\
&\hspace{3cm}
-\left. \sinh{x\frac{f_2}{2}}\cosh{x\frac{f_1}{2}}\sinh{\frac{f_p}{2}}\sinh{\frac{f_q}{2}}(\delta_{1\bar{p}}\tilde{B}^\dagger_{\bar{q}2} + \delta_{1\bar{q}}\tilde{B}^\dagger_{\bar{p}2} + \delta_{2\bar{p}}\tilde{B}^\dagger_{\bar{q}1} + \delta_{2\bar{q}}\tilde{B}^\dagger_{\bar{p}1}) \right] \nonumber \\
&
+\frac{2}{\omega_1-\omega_2+\omega_3+\omega_4}\left(\cosh{x\frac{f_2}{2}}\sinh{x\frac{f_1}{2}}\tilde{B}_{\bar{2}1}+\sinh{x\frac{f_2}{2}}\cosh{x\frac{f_1}{2}}B^\dagger_{\bar{2}1}\right)\times \nonumber \\
&
\left[\cosh{x\frac{f_3}{2}}\cosh{x\frac{f_4}{2}}\sinh{\frac{f_p}{2}}\sinh{\frac{f_q}{2}}(\delta_{4p}\tilde{B}_{q3} + \delta_{4q}\tilde{B}_{p3}) - \sinh{x\frac{f_3}{2}}\sinh{x\frac{f_4}{2}}\cosh{\frac{f_p}{2}}\cosh{\frac{f_q}{2}}(\delta_{4p}B^\dagger_{q3} + \delta_{4q}B^\dagger_{p3}) \right] \nonumber \\
&
+\frac{2}{-\omega_1+\omega_2+\omega_3+\omega_4}\left(\cosh{x\frac{f_2}{2}}\sinh{x\frac{f_1}{2}}\tilde{B}^\dagger_{2\bar{1}}+\sinh{x\frac{f_2}{2}}\cosh{x\frac{f_1}{2}}B_{2\bar{1}}\right)\times \nonumber \\
&
\left[\cosh{x\frac{f_3}{2}}\cosh{x\frac{f_4}{2}}\sinh{\frac{f_p}{2}}\sinh{\frac{f_q}{2}}(\delta_{4\bar{p}}\tilde{B}_{q3} + \delta_{4\bar{q}}\tilde{B}_{p3}) - \sinh{x\frac{f_3}{2}}\sinh{x\frac{f_4}{2}}\cosh{\frac{f_p}{2}}\cosh{\frac{f_q}{2}}(\delta_{4p}B^\dagger_{q3} + \delta_{4q}B^\dagger_{p3}) \right] \nonumber \\
&
+\frac{2}{\omega_1+\omega_2-\omega_3+\omega_4}\left(\cosh{x\frac{f_3}{2}}\cosh{x\frac{f_4}{2}}\tilde{B}_{\bar{3}4}+\sinh{x\frac{f_3}{2}}\sinh{x\frac{f_4}{2}}B^\dagger_{\bar{3}4}\right)\times \nonumber \\
&\hspace{3cm}
\left[+\cosh{x\frac{f_2}{2}}\sinh{x\frac{f_1}{2}}\sinh{\frac{f_p}{2}}\sinh{\frac{f_q}{2}}(\delta_{1\bar{p}}\tilde{B}_{q2} + \delta_{1q}\tilde{B}_{p2}+\delta_{2p}\tilde{B}_{q1} + \delta_{2q}\tilde{B}_{p1})\right. \nonumber \\
&\hspace{3cm}
-\left.\left. \sinh{x\frac{f_2}{2}}\cosh{x\frac{f_1}{2}}\cosh{\frac{f_p}{2}}\cosh{\frac{f_q}{2}}(\delta_{1\bar{p}}B^\dagger_{q2} + \delta_{1q}B^\dagger_{p2} + \delta_{2p}B^\dagger_{q1} + \delta_{2q}B^\dagger_{p1}) \right]\right\}+\left\{\dagger\right\} .
\end{align}
\end{small}

After plugging in the zero temperature ground state values
\begin{equation}
    \langle B(\vec{k}_1,\vec{k}_2)\rangle=\langle \widetilde{B}(\vec{k}_1,\vec{k}_2)\rangle=\langle B^\dagger(\vec{k}_1,\vec{k}_2)\rangle=\langle \widetilde{B}^\dagger(\vec{k}_1,\vec{k}_2)\rangle=\frac{1}{2}(2\pi)^d\delta^d (\vec{k}_1-\vec{k}_2),
\end{equation}
the expression drastically simplifies:
\begin{align}
        &-c\int_0^1 \dd{x} [i\hat{G}_4(x), \Phi_{11}(p,q;1)] \nonumber \\
        = &
        -\frac{c \, \delta^d(p-q)}{4\omega_p\omega_q(\omega_p+\omega_q)}\int_0^1 \dd{x} \frac{\dd[d]{k_1}}{(2\pi)^d}\frac{f_1}{2\omega_1}\sinh{x f_1}\times
        \nonumber \\
        &
        \left\{\cosh{\frac{xf_p}{2}}\cosh{\frac{xf_q}{2}}\cosh{\frac{f_p}{2}}\cosh{\frac{f_q}{2}}-\sinh{\frac{xf_p}{2}}\sinh{\frac{xf_q}{2}}\sinh{\frac{f_p}{2}}\sinh{\frac{f_q}{2}}\right. \nonumber \\
        &
        \left.+\cosh{\frac{xf_p}{2}}\cosh{\frac{xf_q}{2}}\sinh{\frac{f_p}{2}}\sinh{\frac{f_q}{2}}-\sinh{\frac{xf_p}{2}}\sinh{\frac{xf_q}{2}}\cosh{\frac{f_p}{2}}\cosh{\frac{f_q}{2}}\right\} .
\end{align}
Then, we can perform the integral,
\begin{equation}
-c\int_0^1 \dd{x}\langle[i\hat{G}_4(x),\Phi_{11}(1)]\rangle = -c\delta^d(p-q)\frac{1}{4\omega_p^3}\cosh{2\theta_p}\int \frac{\dd[d]{k_1}}{(2\pi)^d}\frac{\sinh^2{(f_1/2)}}{\omega_1}
\end{equation}
\begin{equation}
\langle e^{-i\hat{G}_2}\Phi_{11}(p,q)e^{i\hat{G}_2}\rangle = \delta^d(p-q)\frac{\cosh{f_p}}{2\omega_p}-c\delta^d(p-q)\frac{\cosh{f_p}}{8\omega_p^3}\int \frac{\dd[d]{k_1}}{(2\pi)^d}\frac{1}{\omega_1} + \mathcal{O}(c^2).
\end{equation}
Putting things together, the symmetry transformed configuration for $\Phi_{11}$ is given by
\begin{equation}
\label{eqn:phi11}
\langle\Phi_{11}^f(p,q)\rangle = \delta^d(p-q)\frac{\cosh{f_p}}{2\omega_p}-c \, \delta^d(p-q)\frac{\cosh{f_p}}{8\omega_p^3}\int \frac{\dd[d]{k_1}}{(2\pi)^d}\frac{\cosh{f_1}}{\omega_1} + \mathcal{O}(c^2).
\end{equation}

Its meaning is to transform a  non-thermal ground state solution into a thermal one. This will be seen fully in the framework of the large $N$ collective theory in the next section.

In summary, we have, based on a perturbative construction, established the existence of a nontrivial dynamical symmetry in the case of interacting QFT. The form and construction of the symmetry generator appear complex. We have seen an indication, however, that both the construction and the full expression for the symmetry are simplified with a large $N$ limit. Indeed, in the following section we present the form and existence of the full symmetry through large $N$ in a complete and simple form. In parallel, we mention the existence of a similar symmetry in the only known example of the SYK model. In this case the form and construction \cite{Das:2020kmt, Kitaev:2017awl} of the symmetry again take a rather complex form at nontrivial coupling $J$, while it is simplified, taking the Schwarzian form at large $J$ and $N$.

\section{Collective Theory}
\label{sec:collective-field-theory}
In this section, we study the TFD dynamics in the framework of collective field theory \cite{Das:2003vw} appropriate at large $N$. The Hamiltonian for the interacting $O(N)$ vector model \eqref{eq:H_interacting} can also be written as
\begin{equation}
    H = \int \left[\frac{1}{2} \pi^2 + \frac{1}{2} (\nabla \varphi)^2 + \frac{1}{2}(m^2 + \sigma)\varphi^2 - \frac{N}{4c}\sigma^2 \right] \mathrm{d}^{d}x,
\end{equation}
and similarly for the double $\widetilde{H}$.
%
%
As we have emphasized, our study features the real-time Hamiltonian $\hat{H} = H -  \widetilde{H}$. The collective representation (in terms of bi-local fields) for this system was given in \cite{Jevicki:2015sla}, where the large $N$ limit was studied for a free case. It is our purpose in this section to extend the findings to the interacting theory. Denoting the bi-local collective fields 
\begin{equation} \label{eq:bi-local-def}
    \Phi(\vec{x}, \vec{y}) \,\equiv\, 
    \frac{1}{N}
    \begin{pmatrix}
        \Phi_{11}(\vec{x}, \vec{y}) & \Phi_{12}(\vec{x}, \vec{y}) \\
        \Phi_{21}(\vec{x}, \vec{y}) & \Phi_{22}(\vec{x}, \vec{y})
    \end{pmatrix}
    \,=\,
    \frac{1}{N}
    \begin{pmatrix}
        \varphi(\vec{x}) \cdot \varphi(\vec{y}) & i \varphi(\vec{x}) \cdot \widetilde{\varphi}(\vec y) \\
        i \widetilde{\varphi}(\vec x) \cdot \varphi(\vec{y}) & - \widetilde{\varphi}(\vec x) \cdot \widetilde{\varphi}(\vec y)
    \end{pmatrix},
\end{equation}
where we multiply each tilde vector field with an imaginary unit $i$, which as explained in \cite{Jevicki:2015sla}, allows for an elegant description of the effective Hamiltonian. We also introduce a diagonal matrix $\Sigma$ 
\begin{equation}
    \label{eq:sigma-Phi}
    \Sigma(\vec{x}) \,\equiv\, 
    \begin{pmatrix}
        \sigma(\vec{x}) & 0                           \\
        0                   & \tilde{\sigma}(\vec{x})
    \end{pmatrix}
    \,=\,
    \begin{pmatrix}
        c \: \phi_{11}(\vec{x}, \vec{x}) & 0                                 \\
        0                                & c \: \phi_{22}(\vec{x}, \vec{x})
    \end{pmatrix}.
\end{equation}
The thermofield collective Hamiltonian then takes the form 
\begin{align}
    \hat{H}_{\operatorname{coll}}\, =\,& \frac{2}{N} \Tr(\Pi \star \Phi \star \Pi) + \frac{N}{8} \Tr(\Phi^{-1}) + \frac{N}{2} \Tr((-\nabla^2 + m^2 + \Sigma) \star \Phi)\nonumber \\
    &- \frac{N}{4c}\int (\sigma^2 - \tilde{\sigma}^2) \mathrm{d}^{d}x,
\end{align}
with the star product being understood as
\begin{equation*}
    (A \star B)(\vec{x}, \vec{z}) \equiv \int A(\vec{x}, \vec{y}) B(\vec{y}, \vec{z}) \mathrm{d}^{d} y. 
\end{equation*}
Here, $\Pi$ denotes the conjugate momentum of the bi-local field $\Phi$.

From the saddle point equation
\[
    \frac{\delta \hat{H}_{\operatorname{coll}}}{\delta \Phi} \,=\, 0 ,
\]
one can find a stationary background solution. Explicitly, the saddle point equation can be reduced to
\begin{equation}
    \label{eq:background-field-eq}
    \frac{1}{2} \Phi \star (-\nabla^2 + m^2 + \Sigma) \star \Phi = \frac{1}{8} \mathbb{I},
\end{equation}
where $\mathbb{I}$ is the identity of the bi-local collective space.

This equation, first of all, has a (decoupled) vacuum solution, where each of the diagonal collective fields represents the ground state solution of the respective Hamiltonians. Regarding the more general solution, it was found in \cite{Jevicki:2015sla} that one has a (one-parameter) set of solutions (the free parameter being $f(k)$). The parameter was seen in \cite{Jevicki:2015sla} to have an identification in terms of the temperature. In view of the (dynamical) symmetry of the previous section, we now expect that a one-parameter set of solutions appears generally.

We use a subscript $f$ to denote the expected $f$ dependence. This is given as follows. Due to exchange symmetry (between tilde and non-tilde fields), we expect
\begin{equation}
    \sigma_f (\vec{x}) \,=\, \tilde{\sigma}_f (\vec{x}).
\end{equation}
Then, the saddle point Equation \eqref{eq:background-field-eq} can be considered in momentum space. An ansatz for the solution reads 
\begin{equation}
    \label{eq:Phi_f}
    \Phi_{f} (\vec{x}, \vec{y}) = \int \frac{\mathrm{d}^{d} k}{(2\pi)^{d}}\frac{1}{2 \omega_f(\vec{k})}
    \begin{pmatrix}
        \cosh f(\vec{k}) & i \sinh f(\vec{k}) \\
        i \sinh f(\vec{k}) & - \cosh f(\vec{k})
    \end{pmatrix} 
    e^{i \vec{k} \cdot (\vec{x} - \vec{y})},
\end{equation}
where we have taken the translational invariance and the symmetry of the bi-local collective field into account, and likewise, for the $f$-dependent matrix factor. The free parameter $f(\vec{k})$, much like in the free case, will be an arbitrary even function representing the symmetry freedom of the thermal background, as shown in Equation \eqref{eq:background-field-eq}. We see that the $f$ factor in the ansatz takes the same form as in the free case \cite{Jevicki:2015sla}. However, the dispersion relation $\omega_{f}(\vec{k})$ differs, and will be seen to obey an $f$-dependent gap equation. We have
\begin{equation}
    \label{eq:dispersion-f}
    \omega_f^2 (\vec{k}) = \vec{k}^2 + m^2 + \sigma_{f},
\end{equation}
representing the dispersion relation of the thermal background field in the presence of a quartic interaction. Due to the translational invariance, $\sigma_{f}$ is a constant at the thermal background. In the large $N$ limit, the extra term represents the summation of all bubble diagrams. Placing the ansatz into the collective equation, one can see that these are obeyed provided $\sigma_{f}$ obeys the following ($f$-dependent) gap equation
\begin{equation}
    \sigma_{f} = \int \frac{\mathrm{d}^{d} k}{(2\pi)^{d}} \frac{c_f(\vec{k})}{2 \sqrt{\vec{k}^2 + m^2 + \sigma_{f}}},
\end{equation}
where we have the $f$-dependent quartic coupling
\begin{equation}
    c_{f}(\vec{k}) = c \cdot \cosh f(\vec{k}),
\end{equation}
indicating the temperature and momentum dependence.

To summarize, working in the real time formalism, we have exhibited the appearance of a one-parameter class of background solutions. This we attributed to the existence of a dynamical symmetry operating on the real-time portion of the Schwinger--Keldysh contour. We can explicitly see the agreement with the symmetry-generated configuration discussed in the last section. We expand \eqref{eq:Phi_f}, in particular, $1/\omega_f(\vec{p})$ in order $c$:
\begin{equation}
    \frac{1}{\omega_f(\vec{p})} = \frac{1}{\omega(\vec{p})} -\frac{c}{4\omega^3(\vec{p})}\int \frac{\dd^d\vec{k}}{(2\pi)^d}\frac{\cosh{f(\vec{k})}}{\omega(\vec{k})} .
\end{equation}
Substituting this expansion in the collective field, we see that the result matches with Equation \eqref{eqn:phi11}, which was generated by the symmetry argument.

Finally, one can also show that $\hat{H}_{\operatorname{coll}}$ vanishes in the above background:
\begin{equation}
 \hat{H}_{\operatorname{coll}}^{(0)} = \frac{N}{8} \Tr(\Phi_{f}^{-1}) + \frac{N}{2} \Tr((-\nabla^2 + m^2 + \Sigma_{f}) \star \Phi_{f}) - \frac{N}{4c}\int (\sigma_f^2 - \tilde{\sigma}_f^2) \mathrm{d}^{d}x= 0.  
\end{equation}
This means that the thermofield Hamiltonian annihilates the thermo vacuum state
in the leading approximation.

We note that $f(\vec{k})$ can be specified, and seen to be related to the inverse temperature $\beta$, through comparison with the two-point function $\langle \varphi^i(\vec{x}) \varphi^i(\vec{x}) \rangle_{\beta}$ (no summation) at finite temperature. This is given in Appendix \ref{appendix:finite-temperature-review}, where the identification $f(\vec{k}) = 2 \theta_{\beta}(\vec{k})$ can be found. Finally, we mentioned earlier the  similarity with the collective \cite{Jevicki:2016bwu} background seen in the SYK model \cite{Maldacena:2016hyu}, where the symmetry parameter $f(t)$ was also seen to produce temperature dependence, effectively becoming the gravitational mode. However, in the SYK case \cite{Kitaev:2017awl} this symmetry appears at the conformal point \cite{Das:2020kmt}, while in the present case no such specification is needed. The symmetry (and its spontaneous breaking) appears to be a general property of TFD.

\section{Large \texorpdfstring{$N$}{N} Thermofield Double State}
\label{sec:TFD-state}

We will now employ the collective formalism (with the associated symmetry) to discuss a construction of the thermofield double state (TFDS). It will be presented in the wavefunction(al) form, with the dynamical large $N$ collective variables being canonical bi-local fields. Since in general the collective representation is realized through a change in variables through a Jacobian, the corresponding wavefunction will contain this nontrivial contribution. The large $N$ approximation then follows through a stationary shift, and the collective wavefunction in general takes a Gaussian form, representing the thermofield double ground state. Generally, we will present a scheme for the construction of the TFD wavefunction(al) from solutions of the real-time collective wave equation. We will elaborate on various symmetry issues characterizing the solution.

\subsection{Direct Construction}
Let us first discuss the large $N$ (collective) wavefunction in the solvable case of free theory. One can start from the TFD state written on a creation--annihilation basis 
\begin{equation}
\ket{0(\beta)}=Z^{-1/2}\exp [\int g_\beta (\vec{k})a^\dagger_{\vec{k}} \widetilde{a}^\dagger_{\vec{k}} \dd[d]{k}]\ket{0},
\end{equation}
where $Z$ is the canonical partition function and $g_\beta (\vec{k}) = e^{-\beta \omega(\vec{k})/2}$ is the fugacity, with $\omega_{k} \equiv \sqrt{k^2 + m^2}$ being the free theory dispersion relation. Using the coherent state representation
\begin{equation}
\begin{split}
\bra{\varphi,\tilde{\varphi}}=&
(2\pi)^{-d/2}
e^{-\frac{1}{2}\int \omega_{\vec{k}} (\varphi^2(\vec{k})+ \widetilde{\varphi}^2(\vec{k})) \dd[d]{k}}\\
&\times 
\bra{0}
\exp[\int \qty(-\frac{1}{2}a_{\vec{k}}^2 + \sqrt{2\omega_{\vec{k}}}\varphi(\vec{k}) a_{\vec{k}}) \dd[d]{k}]
\exp[\int \qty(-\frac{1}{2}\widetilde{a}_{\vec{k}}^2 + \sqrt{2\omega_\vec{k}}\tilde{\varphi}(\vec{k}) \tilde{a}_{\vec{k}}) \dd[d]{k}],
\end{split}
\end{equation}
one can then  evaluate the field space wavefunction(al). It is convenient to introduce the double notation
\begin{equation}
\mathbf{a}_k=
\begin{pmatrix}
a_{\vec{k}} \\ 
\widetilde{a}_{\vec{k}}
\end{pmatrix}, 
\quad 
\mathbf{a}^\dagger_k=\begin{pmatrix}
a_{\vec{k}}^\dagger \\
\widetilde{a}_{\vec{k}}^\dagger
\end{pmatrix},  
\quad 
\bm{\varphi}(\vec{k})=
\begin{pmatrix}
\varphi(\vec{k}) \\
\widetilde{\varphi}(\vec{k}
\end{pmatrix},
\end{equation}
so we can express the thermofield wavefunction(al) in field space as
\begin{small}
\begin{align}
\Psi_\beta [{\varphi}, \widetilde{{\varphi}}] = & \frac{
    \exp(-\frac{1}{2} \int \omega_{\vec{k}} \vec{\bm{\varphi}}^2 (\vec{k}) \dd[d]k)
    }{(2\pi)^{d/2} Z^{1/2}} 
\bra{0} \nonumber \\
& \times
\exp[\int \mqty(-\frac{1}{2}\vec{\mathbf{a}}_{\vec{k}}\cdot I \cdot \vec{\mathbf{a}}_{\vec{k}}+\sqrt{2\omega_{\vec{k}}}\vec{\bm{\varphi}}(\vec{k})\cdot\vec{\mathbf{a}}_{\vec{k}}) \dd[d]{k}]
\exp[\frac{1}{2}\int g_\beta(\vec{k})\vec{\mathbf{a}}^\dagger_{\vec{k}} \cdot \Gamma \cdot \vec{\mathbf{a}}^\dagger_{\vec{k}} \dd[d]{k}]\ket{0},
\end{align}
\end{small}
where 
\begin{equation}
I = 
\begin{pmatrix}
1 & 0\\
0 & 1
\end{pmatrix}, \quad
\Gamma = 
\begin{pmatrix}
0 & 1 \\
1 & 0
\end{pmatrix}.
\end{equation}
Using the formula
\begin{equation}
\begin{split}
&\bra{0} \exp [\frac{1}{2}a\cdot M_1 \cdot a + L_1 \cdot a] \exp [\frac{1}{2}a^\dagger\cdot M_2 \cdot a^\dagger + L_2 \cdot a^\dagger] \ket{0} \\
&= (\det M)^{-1/2}\exp [L_1\cdot M \cdot L_2 + \frac{1}{2}L_1\cdot M^{-1}\cdot M_2 \cdot L_1 + \frac{1}{2}L_2 \cdot M_1 \cdot M^{-1}\cdot L_2]
\end{split}
\end{equation}
with $M= I-M_2\cdot M_1$, we can further simplify the form of the thermofield wavefunction to
\begin{equation}
\Psi_\beta [\varphi, \widetilde{\varphi}] = 
(2\pi)^{-d/2}(Z \operatorname{Det}M)^{-1/2} 
\exp[-\frac{1}{2}\int \omega_{\vec{k}} \: \bm{\vec{\varphi}}(\vec{k})\cdot I^\beta (\vec{k})\cdot \bm{\vec{\varphi}}(\vec{k}) \dd[d]{k}],
\end{equation}
where we have defined
\begin{equation}
I^\beta (\vec{k}) = \frac{I-g_\beta(\vec{k})\Gamma}{I+g_\beta(\vec{k})\Gamma}, \quad M(\vec{k}) = I+g_\beta(\vec{k})\Gamma.
\end{equation}
The above expression can also be recast in terms of $\theta_\beta (\vec{k})= \arctanh(e^{-\beta \omega_{\vec{k}}/2})$
\begin{equation}
I^\beta (\vec{k}) = \cosh (f (\vec{k}))I - \sinh (f (\vec{k}))\Gamma, \quad \operatorname{Det} M = \int \frac{\dd[d]{k}}{\cosh^2(f (\vec{k})/2)} .
\end{equation}

As a check, we find that in the high-temperature limit $\beta\rightarrow 0$, $f$ becomes independent of the momentum $\vec{k}$,
\begin{align}
\Psi_0[\varphi, \widetilde{\varphi}] & = \lim_{f\rightarrow \infty} (2\pi Z)^{-1/2}\frac{1}{2}e^{f/2} 
\exp[-\frac{1}{4}e^{f}\int \omega_{\vec{k}}(\varphi(\vec{k}) - \widetilde{\varphi}(\vec{k}))^2\dd[d]{k}]\\
&= (2 Z)^{-1/2}\lim_{l\rightarrow 0}\frac{1}{l\sqrt{\pi}}
\exp[-\frac{1}{l^2}\int \omega_k (\varphi(\vec{k})-\tilde{\varphi}(\vec{k}))^2 \dd[d]{k}]\\
&= (2 Z)^{-1/2}\delta \left(\sqrt{\int\omega_{\vec{k}} (\varphi(\vec{k})-\widetilde{\varphi}(\vec{k}))^2 \dd[d]{k}}\right),
\end{align}
which is exactly the maximally entangled state wavefunction.

The associated collective field wavefunction also contains a Jacobian prefactor \cite{Andric:1985ck} and reads
\begin{equation}
\begin{split}
\Psi_\beta[\Phi] &\sim J^{1/2}[\Phi]
\exp[-\frac{N}{2}\sum_{a,b=1}^2\int \omega_{\vec{k}} \delta^{d}(\vec{k}-\vec{p}) \tilde{I}^\beta_{ab}(\vec{k})\Phi_{ab}(\vec{k},\vec{p}) \dd[d]{k}\dd[d]{p}] \\
&\sim \exp[-A[\Phi]] ,
\end{split}
\end{equation}
with
\begin{equation}
    \tilde{I}^\beta(\vec{k})=
    \begin{pmatrix}
    \cosh{f}(\vec{k}) & i\sinh{f}(\vec{k})\\
    i\sinh{f}(\vec{k}) & -\cosh{f}(\vec{k})
    \end{pmatrix}.
\end{equation}
We wrote the wavefunction(al) in exponential form with an `action' $A$ to prepare for the stationary phase approximation to be used. A subsequent $1/N$ expansion will lead to the large $N$ form of the TFD wavefunction(al).

Using the Jacobian $J[\Phi] = (\operatorname{Det}[\Phi])^{N/2}$, we have the action in the exponent as
\begin{equation}
-A = \frac{N}{4}\Tr\log\Phi 
    - \frac{N}{2} \sum_{a,b=1}^2\int \omega_{\vec{k}} \delta^{d}(\vec{k}-\vec{p})\tilde{I}^\beta_{ab}(\vec{k})\Phi_{ab}(\vec{k},\vec{p}) \dd[d]{k} \dd[d]{p}.
\end{equation}
The large $N$ saddle point equation reads
\begin{equation}
-(\Phi^{-1})_{ba}(\vec{k},\vec{p})+2\omega_{\vec{k}}\:\delta^{d}(\vec{k}-\vec{p})\tilde{I}^\beta_{ab}(\vec{k}) = 0,
\end{equation}
and is solved by the bi-local collective field of the form
\begin{equation}
\Phi^\beta_{ab}(\vec{k},\vec{p})=\frac{\delta^{d}(\vec{k}-\vec{p})}{2\omega_{\vec{k}}}
\begin{pmatrix}
\cosh f(\vec{k}) & i\sinh f(\vec{k})\\
i\sinh f(\vec{k}) & -\cosh f(\vec{k})
\end{pmatrix}.
\end{equation}
We recognize this as the classical stationary point of the collective field seen in Section~\ref{sec:collective-field-theory}. This agreement represents a consistency check of the general form of the collective wavefunction(al).

Next, we perturb around the thermal background $\Phi_{ab}=\Phi_{ab}^\beta + \frac{1}{\sqrt{N}}\eta_{ab}$\footnote{We include the imaginary unit `$i$' and the minus sign `$-$' in the definition of $\eta$ \cite{Jevicki:2015sla}. We also include the complex conjugate numerical factors in the definition of $\pi$ so as to guarantee canonical commutation relations.}, expanding in $1/N$,
\begin{equation}
A = N A_0 + A_2 + \cdots.
\end{equation}
Explicitly,
\begin{equation}
\begin{split}
A_0 &=-\frac{N}{2}\Tr\log\Phi^\beta, \\
A_2 &= \frac{1}{8}\Tr [\Phi_\beta^{-1}\star\eta\star\Phi_\beta^{-1}\star\eta]\\
&=\frac{1}{2}\int \dd[d]{k_1} \dd[d]{k_2} \omega_1\omega_2 
\begin{pmatrix}
\cosh (f(k_1)) & i\sinh (f(k_1))\\
i\sinh (f(k_1)) & -\cosh (f(k_1))
\end{pmatrix}
\begin{pmatrix}
\eta_{11}(\vec{k}_1,\vec{k}_2) & i\:\eta_{12}(\vec{k}_1,\vec{k}_2) \\
i\:\eta_{21}(\vec{k}_1,\vec{k}_2) & -\eta_{22}(\vec{k}_1,\vec{k}_2)
\end{pmatrix}\\
&\, \, \, \, \times
\begin{pmatrix}
\cosh (f(k_2)) & i\: \sinh (f(k_2))\\
i\: \sinh (f(k_2)) & -\cosh (f(k_2))
\end{pmatrix}
\begin{pmatrix}
\ \eta_{11}(\vec{k}_2,\vec{k}_1) & i\eta_{12}(\vec{k}_2,\vec{k}_1) \\
\ i\eta_{21}(\vec{k}_2,\vec{k}_1) & -\eta_{22}(\vec{k}_2,\vec{k}_1)
\end{pmatrix}\\
&=\frac{1}{2}\int \vec{\eta}(\vec{k}_1,\vec{k}_2) \cdot \mathcal{G}_0^{-1}(\vec{k}_1, \vec{k}_2) \cdot \vec{\eta}(\vec{k}_1,\vec{k}_2)
\frac{\dd[d]{k_1}}{(2\pi)^{d}} \, \frac{\dd[d]{k_2}}{(2\pi)^{d}},
\end{split}
\end{equation}
where we use $\theta_{a}\equiv \theta_{\beta}(\vec{k}_{a})$ for convenience. The large $N$ TFD wavefunction(al) therefore takes the Gaussian form (up to an irrelevant normalization factor):
\begin{equation}
\Psi_\beta[\eta] =
\exp[-\frac{1}{2}
        \int  
        \vec{\eta}(\vec{k}_1,\vec{k}_2) \cdot \mathcal{G}_0^{-1}(\vec{k}_1, \vec{k}_2) \cdot \vec{\eta}(\vec{k}_1,\vec{k}_2)
        \, \frac{\dd[d]{k_1}}{(2\pi)^{d}} \frac{\dd[d]{k_2}}{(2\pi)^{d}}],
\end{equation}
where
\begin{equation}
\vec{\eta}(\vec{k}_1,\vec{k}_2) = 
\begin{pmatrix}
\eta_{11}(\vec{k}_1,\vec{k}_2)\\
\eta_{12}(\vec{k}_1,\vec{k}_2)\\
\eta_{21}(\vec{k}_1,\vec{k}_2)\\
\eta_{22}(\vec{k}_1,\vec{k}_2)
\end{pmatrix}
\end{equation}
and
\begin{equation}
\label{eqn:freewf}
\mathcal{G}_0^{-1}(k_1,k_2)=\omega_1^2\omega_2^2
\begin{pmatrix}
c_1 c_2 & -c_1 s_2 & -s_1 c_2 & s_1 s_2 \\
-c_1 s_2 & c_1 c_2 & s_1 s_2 & -s_1 c_2 \\
-s_1 c_2 & s_1 s_2 & c_1 c_2 & -c_1 s_2 \\
s_1 s_2 & -s_1 c_2 & -c_1 s_2 & c_1 c_2 
\end{pmatrix}.
\end{equation}
Here, we denote: $c_p=\cosh{f_p}/\omega_p$ and $s_p=\sinh{f_p}/\omega_p$.

To recapitulate, generally, at large $N$, the ground state or the TFDS wavefunction(al) takes a Gaussian form. The nontrivial information is contained in the associated Green's function. One then has a systematic $1/N$ expansion, which produces higher (polynomial) contributions. We will leave the discussion of these higher corrections to the future. In particular, the wavefunction in one dimension can be evaluated analytically. The relevant calculations are carried out in Appendix \ref{appendix:wavefunction_in_1d}.

\subsection{Collective Construction}

We now proceed to the main topic of this section, namely the construction of the (interacting) TFD wavefunction(al) in the collective scheme. The procedure will be based on the collective representation of the real-time Hamiltonian $\hat{H}$ considered in the previous section. First, shifting the bi-local fields 
\begin{equation}
    \Pi = \sqrt{N} \pi, \quad \Phi = \Phi_{f} + \frac{1}{\sqrt{N}} \eta,
\end{equation}
where the quantum fluctuation $\eta$ and its canonical momentum are defined as vectors
\begin{equation}
\vec{\pi}(k_1,k_2)=
\begin{pmatrix}
\pi_{11}(k_1,k_2)\\
\pi_{12}(k_1,k_2)\\
\pi_{21}(k_1,k_2)\\
\pi_{22}(k_1,k_2)
\end{pmatrix},
\;  \vec{\eta}(k_1,k_2)=
\begin{pmatrix}
\eta_{11}(k_1,k_2)\\
\eta_{12}(k_1,k_2)\\
\eta_{21}(k_1,k_2)\\
\eta_{22}(k_1,k_2)
\end{pmatrix}.
\end{equation}
In the large $N$ limit, then, the wavefunction(al) will be a zero energy eigenfunction corresponding to the quadratic Hamiltonian
\begin{equation} \label{eq:H_coll_quadratic}
    \hat{H}_{\operatorname{coll}}^{(2)} =
    \frac{1}{2} \pi \star K \star \pi + \frac{1}{2} \eta \star \widetilde{V} \star \eta.
\end{equation}
Here, the star product is understood in momentum space as
\begin{equation*}
    (K \star \eta) (k_1, k_2)\equiv \int K(k_1, k_2; k_3, k_4) \cdot \eta(k_3, k_4) \frac{\dd^d k_3}{(2\pi)^d} \frac{\dd^d k_4}{(2\pi)^d},
\end{equation*}
where the symbol `$\cdot$' indicates matrix multiplications. The kinetic matrix $K(k_1, k_2; k_3, k_4)$ is diagonal in the sense that $(k_1, k_2) = (k_3, k_4)$ \cite{Jevicki:2015sla},
\begin{equation}
K(k_1,k_2; k_3, k_4) = 
(2\pi)^{d}\delta^{d}(k_1 - k_3)
(2\pi)^{d}\delta^{d}(k_2 - k_4)
\mathcal{K}(k_1, k_2),
\end{equation}
where
\begin{equation}
\mathcal{K}(k_1, k_2) =
\begin{pmatrix}
c_1+c_2 & s_2 & s_1 & 0\\
s_2 & -c_1+c_2 & 0 & -s_1 \\
s_1 & 0 & c_1-c_2 & -s_2\\
0 & -s_1 & -s_2 & -c_1-c_2
\end{pmatrix}.
\end{equation}
Meanwhile, the potential matrix $\widetilde{V}(k_1, k_2; k_3, k_4)$ is given by a sum
\begin{equation}
    \widetilde{V}(k_1, k_2; k_3, k_4) \equiv V(k_1, k_2; k_3, k_4) + \frac{c}{2} \Delta(k_1, k_2; k_3, k_4),
\end{equation}
where the `free' part $V(k_1, k_2; k_3, k_4)$ is given by
\begin{equation}
    \widetilde{V}(k_1, k_2; k_3, k_4) \equiv (2\pi)^{d}\delta^{d}(k_1 - k_3)
    (2\pi)^{d}\delta^{d}(k_2 - k_4) \mathcal{V}(k_1, k_2),
\end{equation}
with
\begin{equation}
\mathcal{V}(k_1,k_2) = 
\omega_1^2\omega_2^2
\begin{pmatrix}
c_1+c_2 & -s_2 & -s_1 & 0\\
-s_2 & -c_1+c_2 & 0 & s_1 \\
-s_1 & 0 & c_1-c_2 & s_2\\
0 & s_1 & s_2 & -c_1-c_2
\end{pmatrix}.
\end{equation}
Finally, the interaction part is
\begin{equation}
\Delta(k_1,k_2;k_3,k_4) =(2\pi)^d\delta^d (k_1-k_2+k_3-k_4)
\begin{pmatrix}
1 & & & \\
& 0 & & \\
& & 0 & \\
& & & -1
\end{pmatrix}.
\end{equation}

Based on the large $N$ Hamiltonian we now have that the (interacting) thermal ground state wavefunction(al) is represented by a Gaussian form
\begin{equation}
    \Psi_{\operatorname{gs}}[\eta] = \frac{1}{\sqrt{Z}} 
    \exp(- \frac{1}{2} \eta \star G^{-1} \star \eta).
\end{equation}
which explicitly reads as
\begin{equation}
\Psi_{\operatorname{gs}}[\eta] = \frac{1}{\sqrt{Z}}
\exp\left[
    -\frac{1}{2}\int \eta(k_1,k_2)\cdot G^{-1}(k_1,k_2;k_3,k_4)\cdot\eta(k_3,k_4)\,
    \prod_{i=1}^{4}\frac{\dd[d]{k}_i}{(2\pi)^d}
    \right].
\end{equation}
with a kernel given by a four-point Green's function. In canonical representation, with
\begin{equation}
    \pi_{ab}(k_1, k_2) = -i \frac{\delta}{\delta \eta_{ab}(k_1, k_2)},
\end{equation}
we have the action of the Hamiltonian on this wavefunction as
\begin{equation}
    \hat{H}_{\operatorname{coll}}^{(2)} \Psi_{\operatorname{gs}} = 
    \left[
    \frac{1}{2} \Tr(K \star G^{-1}) + 
    \frac{1}{2}\eta \star 
    \left(G^{-1} \star K \star G^{-1} - \widetilde{V}\right) \star \eta 
    \right] \Psi_{\operatorname{gs}}.
\end{equation}
Requiring the thermal vacuum state to be annihilated by the collective Hamiltonian, we are led to the following equations involving the unknown Green's function $G$:
\begin{equation} \label{eq:TrKG}
    E_1 = \frac{1}{2} \Tr(K \star G^{-1}) = 0,
\end{equation}
and
\begin{equation} \label{eq:GKG-Vtilde}
    G^{-1} \star K \star G^{-1} - \widetilde{V} = 0.
\end{equation}
We see that the (order 1) ground state energy $E_{1}$ will indeed vanishes at the stationary point in this construction (due to various anti-symmetry properties of the solution).

As a check, we turn off the interaction and verify that the nonlinear (quadratic) equation that we have obtained is satisfied by the known (operator method) solution of the free collective theory \eqref{eqn:freewf}. In that case, our solution for $\mathcal{G}_0$ satisfies
\begin{equation}
\label{eqn:free}
-\mathcal{G}_0^{-1}(k_1,k_2)\cdot \mathcal{K}(k_1,k_2)\cdot \mathcal{G}_0^{-1}(k_1,k_2)+\mathcal{V}(k_1,k_2) = 0,
\end{equation}
where translational symmetry was used:
\begin{equation}
G_{0}^{-1}(k_1,k_2;k_3,k_4) = (2\pi)^{d}\delta^d(k_1-k_3)(2\pi)^{d}(k_2-k_4)\mathcal{G}_0^{-1}(k_1,k_2).
\end{equation}
We can verify that the explicit collective solution \eqref{eqn:freewf} obeys our equation. One can check that the ground state energy vanishes through \eqref{eq:TrKG}.

To summarize, in the above, we presented equations that constrain the TFD state wavefunction(al), generally at large $N$ in collective field theory; in addition, to the thermal gap Equation \eqref{eq:GKG-Vtilde}, we now have the quadratic matrix equation for Green's function $G$ defining the large $N$ wavefunction. The quadratic equation for $G$ can also be written in an equivalent form, with the roles of $K$ and $V$ exchanged:
\begin{equation}\label{eq:K-GVtildeG}
    K - G \star \widetilde{V} \star G = 0.
\end{equation}
For understanding the structure of this Equation (and of its thermal solution), we write $G=G_0 +G_1$, where $G_0$ is the solution of the free large $N$ problem (given above) and $G_1$ is the interacting completion. Assuming that $G_0$ is known and obeys Equations \eqref{eq:TrKG} and \eqref{eq:GKG-Vtilde} (or equivalently, Equation \eqref{eq:K-GVtildeG}), we then have the following equations for $G_1$:
\begin{equation}
    \Tr(K \star (G_0 + G_1)^{-1}) = 0,
\end{equation}
\begin{equation}
   \frac{c}{2} G_0 \star \Delta \star G_0 + G_0 \star \widetilde{V} \star G_1 + G_1 \star \widetilde{V} \star G_0
   + G_1 \star \widetilde{V} \star G_1 = 0.
\end{equation}
We saw in the operator construction that the free Green's function takes the form:
\begin{equation}
    G_0(k_1, k_2; k_3, k_4) = 
    (2\pi)^{d}\delta^{d}(k_1 - k_3) (2\pi)^d(k_2 - k_4) \mathcal{G}_0(k_1, k_2),
\end{equation}
while for $G_1$ one can generally write:
\begin{equation}
    G_1(k_1, k_2; k_3, k_4) =
    (2\pi)^{d}(k_1 - k_2 + k_3 - k_4) \mathcal{G}_1(k_1, k_2; k_3, k_4),
\end{equation}
which corresponds to momentum conservation. These delta function structures hold generally for four-point functions at large $N$; see \cite{deMelloKoch:1996mj}.

We will now discuss the symmetry freedom involved in the construction of a solution for~$G$. This is generally related to the existence of zero modes of the kinetic matrix $K$ and the potential matrix $V$ (or $\widetilde{V} = V + c \: \Delta / 2$ in interacting theories). As it was seen in \cite{Jevicki:2015sla}, in the free case, the matrix  $K = K(k_1, k_2; k_3, k_4)$ defining the kinetic term is not invertible at $|k_1| = |k_2|$. This translates into a sequence of momentum-dependent zero modes $u(k_1,k_2 \mid l)$
\begin{equation}
    K \star u = 0
\end{equation}
labeled by the norm of a momentum $l$, i.e., $\abs{l}$. Explicitly,
\begin{equation}
    u(k_1, k_2 \mid l) =  
    \delta(\abs{k_1} - \abs{l})\frac{(2\pi)^d\delta^{d}(k_1 - k_2)}{\sqrt{2+2\coth(f(k_1))}}
    \begin{pmatrix}
    1 \\
    - \coth(f(k_1)) \\
    - \coth(f(k_1)) \\
    1
    \end{pmatrix}.
\end{equation}
This implies that Equations \eqref{eq:TrKG} and \eqref{eq:GKG-Vtilde} allow a family of solutions related by
\begin{equation}
    G^{\prime -1} = G^{-1} + \int_{0}^{\infty} a(\abs{l}, |l^{\prime}|) \: u \otimes u (\abs{l}, |l^{\prime}|) \: \dd \abs{l}\, \dd |l^{\prime}|,
\end{equation}
for any real function $a(|l|,|l^{\prime}|)$. Here, $u \otimes u$ denotes
\begin{equation}
\begin{split}
    & u \otimes u (k_1, k_2; k_3, k_4 \mid l, l^{\prime})\\
    = & \delta(\abs{k_1} - \abs{l}) \delta(\abs{k_3} - |l^{\prime}|)
    \frac{(2\pi)^d\delta^{d}(k_1 - k_2)}{\sqrt{2+2\coth(f(k_1))}}
    \frac{(2\pi)^d\delta^{d}(k_3 - k_4)}{\sqrt{2+2\coth(f(k_3))}} \\
    & \times
    \begin{pmatrix}
     1 & -\coth(f(k_3)) & -\coth(f(k_3)) & 1 \\
     -\coth(f(k_1)) & \coth(f(k_1))\coth(f(k_3)) & \coth(f(k_1))\coth(f(k_3)) & -\coth(f(k_1)) \\
     -\coth(f(k_1)) & \coth(f(k_1))\coth(f(k_3)) & \coth(f(k_1))\coth(f(k_3)) & -\coth(f(k_1)) \\
     1 & -\coth(f(k_3)) & -\coth(f(k_3)) & 1 
    \end{pmatrix}.
\end{split}
\end{equation}
This contribution, however, should be excluded due to the delta function structures. The $K$, $V$ and $G_0$ matrices are all diagonal in bi-local space in the sense that their delta function structures are $\delta^{d}(k_1 - k_3) \delta^d(k_2 - k_4)$. In addition, the interacting potential term $\Delta$ and the corresponding correction $G_1$ matrix both have a structure $\delta^{d}(k_1 - k_2 + k_3 - k_4)$. However, the delta function structure of matrix $u \otimes u$ reads $\delta^{d}(k_1 - k_2) \delta^{d}(k_3 - k_4)$. This represents internal contractions and hence would only influence the background, and thus should be excluded. Once the free part is correctly specified no such contribution is needed.

Consider now similarly the contribution of zero modes associated with the potential term. Clearly, if the matrix $\widetilde{V}$ possesses zero modes, denoted as $v(k_1,k_2 \mid l)$ such that
\begin{equation}
    \widetilde{V} \star v = 0,
\end{equation}
one has an invariance, and a family of solutions related by
\begin{equation}
    G^{\prime} = G + \int_{0}^{\infty} b(\abs{l},|l^{\prime}|) \: v \otimes v (\abs{l},|l^{\prime}|) \: \dd \abs{l} \, \dd |l^{\prime}|,
\end{equation}
where $b(\abs{l}, |l^{\prime}|)$ is again an arbitrary real function. The explicit form of the zero mode, in this case, can be seen using the Goldstone argument:
\begin{align}
    v(k_1, k_2 \mid l) \equiv & \fdv{\Phi_{f}(k_1, k_2)}{f(l)} \nonumber \\
    = &
    (2\pi)^{d}\delta^{d}(k_1-k_2)\frac{\delta(\abs{k_1}-\abs{l})}{2\omega_{f}(k_1)} 
    \begin{pmatrix}
    \sinh(f(k_1)) \\
    \cosh(f(k_1)) \\
    \cosh(f(k_1)) \\
    \sinh(f(k_1)) \\
    \end{pmatrix} \nonumber \\
    &
    - (2\pi)^{d}\delta^{d}(k_1-k_2) \frac{c \: \abs{l}^{d-1} \sinh(f(l))}{8 \mathcal{N} \omega_{f}(k_1)^3 \omega_{f}(l)}
    \begin{pmatrix}
    \cosh(f(k_1)) \\
    \sinh(f(k_1)) \\
    \sinh(f(k_1)) \\
    \cosh(f(k_1))
    \end{pmatrix}.
\end{align}
The factor $\mathcal{N}$ is given by
\begin{equation}
    \mathcal{N} \equiv 
    \frac{(2\pi)^{d}}{S_{d}}
    \left(
    1 + c \int \frac{\cosh(f(p))}{4\omega_{f}(p)^{3}} \frac{\dd^d p}{(2\pi)^d}
    \right),
\end{equation}
where $S_{d}$ is the area of the $d$-dimensional unit sphere. Again, observing the delta function structure of the contribution, we see that it influences the background only. Therefore, with the same argument as before, these contributions are not needed.

This analysis implies that once the free part of the wavefunction is specified, our equations are sufficient for the determination of the full answer. This can be obtained perturbatively, in the coupling constant $c$ (much like the solution of the gap equation), by numerical methods or possibly exactly. One can also contemplate variational solutions.

In summary, elements of the construction of the large $N$ thermofield double state that we have given apply in any collective QFT with interactions. Higher-order terms in the exponent come in powers of $1/N$ and can be systematically evaluated \cite{deMelloKoch:1996mj}.

\section{Conclusion}
\label{sec:conclusion}

We have in this work discussed some constructive elements of the thermofield double QFT. Working on the real-time portion of the Schwinger--Keldysh contour we discussed a dynamical symmetry, which we argued appears in general in any TFD quantum theory. Perturbation theory is used to establish the generator to leading order (in the coupling constant). The corresponding symmetry transformations are discussed, and are seen to represent a nonlinear extension of the well known Bogoliubov transformations of the free theory. The relevance (and use) of the (dynamical) symmetry is then seen in the collective representation of the theory. For the $O(N)$ theory which serves as the model for consideration, one has the bi-local Hamiltonian representation (on the real-time portion of the SK contour). Here, for the interacting case, the existence of a general (all orders in the coupling constant) symmetry is shown. It was seen to imply a set of background solutions representing the thermal backgrounds of the large $N$ theory. The second part of our construction concerns the construction of the TFD State itself at large $N$. This is carried out at the level of fluctuations, and therefore at $\mathcal{O}(1)$ of collective theory. It is seen that the collective TFD state takes a Gaussian form in this limit, and using this form we derive the corresponding eigen-solution. It is  governed by a quadratic matrix equation (in the collective space) whose general solution we also discuss. The determination of zero modes gives a specification of the arbitrariness of the solution. Altogether, this can be interpreted as providing strong nonlinear constraints on the wavefunction(al), akin to Ward identities. TFD symmetries (of BRST type) have been established \cite{Haehl:2015foa, Crossley:2015evo, Haehl:2016pec} to play a useful role in the hydrodynamical description of AdS/CFT. It will be relevant to study the implications of the present construction in that context, and more broadly questions of entanglement \cite{Das:2020xoa}. It is most relevant to extend the construction to matrix-type models. With the recently developed large $N$ numerical method \cite{Koch:2021yeb}, this appears possible.


\acknowledgments{
We would like to thank Robert de Mello Koch, Sumit Das, Masanori Hanada, Cheng Peng and Jo\~{a}o Rodrigues for discussions and various comments regarding this problem. The work of A.J., X.L. and J.Z. was partially supported by the U.S. Department of Energy under contract DE-SC0010010 and DE-SC0019480. The work of J.Y. is supported by a National Research Foundation of Korea (NRF) grant funded by the Korean government (MSIT) (No. 2019R1F1A1045971). J.Y. is supported by an appointment to the JRG Program at the APCTP through the Science and Technology Promotion Fund and Lottery Fund of the Korean Government, and is also supported by the Korean Local Governments---Gyeongsangbuk-do Province and Pohang City.
}

\appendix

\section{Algebra of Bi-local Operators} 
\label{appendix:bi-local-algebra}

\subsection{Definition}
We define the $O(N)$ invariant bi-local operators as
\begin{align}
  A(\vec{k}_1, \vec{k}_2) & \equiv a(\vec{k}_1) \cdot a(\vec{k}_2),\\
  A^{\dagger}(\vec{k}_1, \vec{k}_2) & \equiv a^{\dagger}(\vec{k}_1) \cdot a^\dagger(\vec{k}_2), \\
  B(\vec{k}_1, \vec{k}_2) & \equiv a^{\dagger}(\vec{k}_1) \cdot a(\vec{k}_2) + \frac{N}{2}\delta^{d}(\vec{k}_1 - \vec{k}_2), \\
  C (\vec{k}_1, \vec{k}_2) &\equiv a(\vec{k}_1) \cdot \widetilde{a}(\vec{k}_2), \\
  D (\vec{k}_1, \vec{k}_2) & \equiv a(\vec{k}_1) \cdot \widetilde{a}^{\dagger}(\vec{k}_2), \\
  C^{\dagger}(\vec{k}_1, \vec{k}_2) & \equiv a^{\dagger}(\vec{k}_1) \cdot \widetilde{a}^{\dagger}(\vec{k}_2), \\ 
  D^{\dagger}(\vec{k}_1, \vec{k}_2) & \equiv a^\dagger(\vec{k}_1) \cdot \widetilde{a}(\vec{k}_2),
\end{align}
and similarly for $\biloc{\AT}{1}{2}$, $\biloc{\ATD}{1}{2}$, and $\biloc{\BT}{1}{2}$. In the definition of $B$ and $\widetilde{B}$ we include central terms so that the following calculations are simplified.

\subsection{The Algebra}
The commutation relations of these bi-local operators can be easily computed using commutators \eqref{eq:commutator-a-a-dagger} and \eqref{eq:commutator-at-at-dagger}, and one can verify that they form a basis of an algebra. It is easy to verify that these operators have the properties
\begin{align}
    \biloc{A}{1}{2} & = \biloc{A}{2}{1}, & \biloc{\AT}{1}{2} & = \biloc{\AT}{2}{1}, \\
    \biloc{B}{1}{2} & = \biloc{B^{\dagger}}{2}{1}, & \biloc{\BT}{1}{2} & = \biloc{{\BTD}}{2}{1}, \\
    \biloc{C}{1}{2} & = \biloc{\widetilde{C}}{2}{1}, & \biloc{\widetilde{D}}{1}{2} & = \biloc{\widetilde{D}^{\dagger}}{2}{1},
\end{align}
and similarly for $\AT$ and $\ATD$. A full listing of the algebra is too lengthy, and here we only present the essential parts. Other commutation relations can also be derived from the above properties.
\begin{align}
    \COMM{\biloc{A}{1}{2}}{\biloc{\AD}{3}{4}} = & \quad \DD{2}{3} \biloc{B}{4}{1} + \DD{2}{4} \biloc{B}{3}{1} \\ \nonumber
                                                & +      \DD{1}{3} \biloc{B}{4}{2} + \DD{1}{4} \biloc{B}{3}{2}. \\
    \COMM{\biloc{A}{1}{2}}{\biloc{B}{3}{4}} = & \quad \DD{2}{3} \biloc{A}{1}{4} - \DD{1}{3} \biloc{A}{4}{2}. \\
    \COMM{\biloc{\AD}{1}{2}}{\biloc{B}{3}{4}} = & - \DD{4}{2} \biloc{\AD}{1}{3} - \DD{4}{1} \biloc{\AD}{3}{2}. \\
    \COMM{\biloc{B}{1}{2}}{\biloc{B}{3}{4}} = & \quad \DD{2}{3} \biloc{B}{1}{4} - \DD{1}{4} \biloc{B}{3}{2}.
\end{align}
\begin{align}
    \COMM{\biloc{A}{1}{2}}{\biloc{C^{\dagger}}{3}{4}} = & \quad \DD{1}{3} \biloc{D}{2}{4} + \DD{2}{3} \biloc{D}{1}{4}. \\
    \COMM{\biloc{\AD}{1}{2}}{\biloc{C}{3}{4}} = & - \DD{3}{2} \biloc{D^{\dagger}}{1}{4} - \DD{3}{1} \biloc{D^{\dagger}}{2}{4}. 
\end{align}
\begin{align}
    \COMM{\biloc{A}{1}{2}}{\biloc{D^{\dagger}}{3}{4}} & = \quad  \DD{3}{1} \biloc{C}{2}{4} + \DD{3}{2} \biloc{C}{1}{4}. \\
    \COMM{\biloc{\AD}{1}{2}}{\biloc{D}{3}{4}} & = - \DD{3}{1} \biloc{C^{\dagger}}{2}{4} - \DD{3}{2} \biloc{C^{\dagger}}{1}{4}.
\end{align}
\begin{align}
    \COMM{\biloc{C}{1}{2}}{\biloc{D}{3}{4}} & = \quad\DD{4}{2} \biloc{A}{1}{3}. \\
    \COMM{\biloc{C^{\dagger}}{1}{2}}{\biloc{D^{\dagger}}{3}{4}} & = - \DD{4}{2} \biloc{\AD}{1}{3}. \\
    \COMM{\biloc{C}{1}{2}}{\biloc{D^{\dagger}}{3}{4}} & = \quad \DD{3}{1} \biloc{\AT}{4}{2}. \\
    \COMM{\biloc{C^{\dagger}}{1}{2}}{\biloc{D}{3}{4}} & = - \DD{3}{1} \biloc{\ATD}{4}{2}.
\end{align}
\begin{align}
    \COMM{\biloc{B}{1}{2}}{\biloc{C}{3}{4}} & = - \DD{1}{3} \biloc{C}{2}{4}. \\
    \COMM{\biloc{B}{1}{2}}{\biloc{C^{\dagger}}{3}{4}} & = \quad \DD{2}{3} \biloc{C^{\dagger}}{1}{4}. \\
    \COMM{\biloc{C}{1}{2}}{\biloc{C^{\dagger}}{3}{4}} & = \quad \DD{2}{4} \biloc{B}{3}{1} + \DD{1}{3} \biloc{\BT}{2}{4}.
\end{align}
\begin{align}
    \COMM{\biloc{B}{1}{2}}{\biloc{D}{3}{4}} & = - \DD{1}{3} \biloc{D}{2}{4}. \\
    \COMM{\biloc{B}{1}{2}}{\biloc{D^{\dagger}}{3}{4}} & = \quad \DD{2}{3} \biloc{D^{\dagger}}{1}{4}. \\
    \COMM{\biloc{D}{1}{2}}{\biloc{D^{\dagger}}{3}{4}} & = - \DD{4}{2} \biloc{B}{3}{1} + \DD{3}{1} \biloc{\BT}{2}{4}.
\end{align}

\subsection{Useful Relations}
For the calculation of $\hat{G}_{4}$ it is convenient to define two linear combinations of the bi-local operators which have already appeared in $\hat{H}_2$ \eqref{eq:H2} and $\hat{G}_2$ \eqref{eq:G2}
\begin{equation}
    \hat{\mathcal{H}}_2 (\vec{k}) = \biloc{B}{}{} - \biloc{\widetilde{B}}{}{},
\end{equation}
\begin{equation}
    \hat{\mathcal{G}}_2 (\vec{k}) = \biloc{C^{\dagger}}{}{} - \biloc{C}{}{}.
\end{equation}
We then present the commutation relations between them and other elementary bi-local operators.

Firstly, the commutators of $\Gc{}$ and other operators are
\begin{align}
    \COMM{\Gc{}}{\biloc{A}{1}{2}} & = -\DD{}{1}\biloc{D}{2}{} - \DD{}{2}\biloc{D}{1}{}, \\
    \COMM{\Gc{}}{\biloc{\AD}{1}{2}} & = -\DD{}{1}\biloc{D^{\dagger}}{2}{} - \DD{}{2}\biloc{D^{\dagger}}{1}{}, \\
    \COMM{\Gc{}}{\biloc{B}{1}{2}} & = -\DD{}{1}\biloc{C}{2}{} - \DD{}{2}\biloc{C^{\dagger}}{1}{}.
\end{align}
\begin{align}
    \COMM{\Gc{}}{\biloc{\AT}{1}{2}} & = -\DD{}{1}\biloc{D^{\dagger}}{}{2} - \DD{}{2}\biloc{D^{\dagger}}{}{1}, \\
    \COMM{\Gc{}}{\biloc{\ATD}{1}{2}} & = -\DD{}{1}\biloc{D}{}{2} - \DD{}{2}\biloc{D}{}{1}, \\
    \COMM{\Gc{}}{\biloc{\BT}{1}{2}} & = -\DD{}{1}\biloc{C}{}{2} - \DD{}{2}\biloc{C^{\dagger}}{}{1}.
\end{align}

Secondly, the commutators of $\Hc{}$ and other operators are
\begin{align}
    \COMM{\Hc{}}{\biloc{A}{1}{2}} & = -\DD{}{1}\biloc{A}{}{2} - \DD{}{2}\biloc{A}{1}{}, \label{eq:COMM-Hc-1} \\
    \COMM{\Hc{}}{\biloc{\ATD}{1}{2}} & = - \DD{}{1}\biloc{\ATD}{}{2} - \DD{}{2}\biloc{\ATD}{1}{}, \\
    \COMM{\Hc{}}{\biloc{D}{1}{2}} & = -\DD{}{1}\biloc{D}{}{2} - \DD{}{2}\biloc{D}{1}{}.
\end{align}
\begin{align}
    \COMM{\Hc{}}{\biloc{\AD}{1}{2}} & = \DD{}{1}\biloc{\AD}{}{2} + \DD{}{2}\biloc{\AD}{1}{}, \\
    \COMM{\Hc{}}{\biloc{\AT}{1}{2}} & = \DD{}{1}\biloc{\AT}{}{2} + \DD{}{2}\biloc{\AT}{1}{}, \\
    \COMM{\Hc{}}{\biloc{D^{\dagger}}{1}{2}} & = \DD{}{1}\biloc{D^{\dagger}}{}{2} + \DD{}{2}\biloc{D^{\dagger}}{1}{}.
\end{align}
\begin{align}
    \COMM{\Hc{}}{\biloc{B}{1}{2}} & = \DD{}{1}\biloc{B}{}{2} - \DD{}{2}\biloc{B}{1}{}, \\
    \COMM{\Hc{}}{\biloc{C^{\dagger}}{1}{2}} & = \DD{}{1}\biloc{C^{\dagger}}{}{2} - \DD{}{2}\biloc{C^{\dagger}}{1}{}.
\end{align}
\begin{align}
    \COMM{\Hc{}}{\biloc{\BT}{1}{2}} & = -\DD{}{1}\biloc{\BT}{}{2} + \DD{}{2}\biloc{\BT}{1}{}, \\
    \COMM{\Hc{}}{\biloc{C}{1}{2}} & = -\DD{}{1}\biloc{C}{}{2} + \DD{}{2}\biloc{C}{1}{}. \label{eq:COMM-Hc-10}
\end{align}

Using these commutation relations, one can easily check that in \eqref{eq:comm-G2-H2} the commutator of $\hat{G}_2$ and $\hat{H}_2$ vanishes pointwise in momentum space
\begin{equation}
    [\hat{G}_2, \hat{H}_2] = \Intk \mathrm{d}^d k^{\prime} i \theta_{\beta}(\vec{k}) \omega(\vec{k}^\prime) [\Gc{}, \mathcal{H}_2(\vec{k^\prime})] = \Intk i \theta_{\beta}(\vec{k}) \omega(\vec{k}) \cdot 0 = 0.
\end{equation}

\section{\texorpdfstring{$O(N)$}{O(N)} Vector Model at Finite Temperature}
\label{appendix:finite-temperature-review}

In this section, we present a brief review of the interacting $O(N)$ vector model at finite temperature. We consider the $O(N)$ vector model with a quartic interaction in $D = d+1$ dimensional Euclidean space--time
\begin{equation}
    \label{eq:S-phi}
    S[\varphi] = \int \left[ \frac{1}{2} (\partial \varphi)^2 + \frac{1}{2} m^2 \varphi^2 + \frac{c}{4 N} (\varphi^2)^2 \right] \mathrm{d}^D x.
\end{equation}
Since we employ the action formalism, we therefore distinguish the $D$ coordinate denoted $x$ with the $d$ coordinate denoted $\vvec{x}$, and similarly for momenta. With an auxiliary field $\sigma(x)$  the action reads
\begin{equation}
    \label{eq:S-phi-sigma}
    S[\varphi, \sigma] = \int \left[\frac{1}{2} (\partial \varphi)^2 + \frac{1}{2} (m^2 + \sigma) \varphi^2 - \frac{N}{4 c} \sigma^2 \right] \mathrm{d}^D x.
\end{equation}
for which one can write an effective action, after integrating  $\varphi(x)$
\begin{equation}
    Z = \int \mathcal{D} \varphi \mathcal{D} \sigma  \exp(- S[\varphi, \sigma]) = \int \mathcal{D} \sigma \exp( - N S_{\operatorname{eff}}[\sigma]),
\end{equation}
where
\begin{equation}
    S_{\operatorname{eff}}[\sigma] = - \frac{1}{4 c} \int \sigma^2(x) \: \mathrm{d}^{D} x + \frac{1}{2} \tr \ln (-\partial^2 + m^2 + \sigma(\cdot)).
\end{equation}
The partition function is then dominated by the saddle point value of the effective action. Let $\sigma_{\beta}$ be the saddle point value of the field $\sigma(x)$. We vary the effective action $S_{\operatorname{eff}}[\sigma]$ at the saddle point
\[
    \left.\frac{\delta S_{\operatorname{eff}}[\sigma] }{\delta \sigma(x)}\right|_{\sigma_{\beta}} = 0
\]
to obtain the `thermal' gap equation
\begin{equation}
    \sigma_{\beta} = c \: \tr[(-\partial^2 + m^2 + \sigma_{\beta})^{-1}].
\end{equation}

With periodic boundary conditions, one has the Matsubara mode expansion
\begin{equation}
    \varphi^j(\tau, \vvec{x}) = \sum_{n\in\mathbb{Z}} \int \frac{\mathrm{d}^{d}k}{(2\pi)^{d}} \: \tilde{\varphi}^j_n(\vvec{k}) \: e^{i \nu_n \tau + i \vvec{k}\cdot \vvec{x}}, \qquad \nu_n = \frac{2 \pi n}{\beta} . 
\end{equation}
Introducing the dispersion relation $\omega_{\beta}(\vvec{k})$ as
\begin{equation}
 \label{eq:hat-omega}
 \omega_{\beta}(\vvec{k}) = \sqrt{\vvec{k}^2 + m^2 + \sigma_{\beta}} = \sqrt{\omega^2(\vvec{k}) + \sigma_{\beta}},    
\end{equation}
we can evaluate the trace term in the gap equation as follows
\begin{align*}
    \tr [(-\partial^2 + m^2 + \sigma_{\beta})^{-1}] & = \sum_{n\in \mathbb{Z}} V_{d} \int\frac{\mathrm{d}^{d} k}{(2\pi)^{d}} \frac{1}{\nu_n^2 + \vvec{k}^2 + m^2 + \sigma_{\beta}} \\
    & = V_{d} \int \frac{\mathrm{d}^{d}k}{(2\pi)^{d}} \frac{\beta}{2\omega_{\beta}(\vvec{k})} \cosh \big(2\theta_{\beta}(\vvec{k})\big),
\end{align*}
where $V_{d}$ is the $d$ dimensional spatial volume. Here, we defined $\theta_{\beta}(\vvec{k})$ as
\begin{equation}
\theta_{\beta}(\vvec{k}) = \arctanh(e^{-\beta \omega_{\beta}(\vvec{k})/2}).    
\end{equation}
We can thus write the gap equation as
\begin{equation}
    \label{eq:sigma-0-gap}
    \sigma_{\beta} = c \beta V_{d} \int \frac{\mathrm{d}^{d}k}{(2\pi)^{d}} 
    \frac{\cosh \big(2\theta_{\beta}(\vvec{k})\big)}{2 \sqrt{\vvec{k}^2 + m^2 + \sigma_{\beta}}}.
\end{equation}

From the equation of motion for $\sigma$, we see that $\sigma_{\beta}$ corresponds to the sum of bubble diagrams in the large $N$ limit, which renormalizes the bare mass. Thus, the two-point function at finite temperature is
\begin{equation}
    \label{eq:2pt-beta}
    \langle \varphi^i(0,\vvec{x}) \varphi^j(0,\vvec{y}) \rangle_{\beta} = \delta^{ij} \int \frac{\mathrm{d}^{d} k}{(2\pi)^{d}} \frac{e^{i\vvec{k}\cdot(\vvec{x} - \vvec{y})}}{2 \omega_{\beta}(\vvec{k})} \cosh 
    \big(2\theta_{\beta}(\vvec{k})\big).
\end{equation}
In the zero temperature limit, $\beta \rightarrow \infty$, hence $\cosh \big(2\theta_{\beta}(\vvec{k})\big) \rightarrow 1$, and we obtain the ordinary large $N$ (equal-time) two-point function.

\section{Thermofield Wavefunction in One Dimension}
\label{appendix:wavefunction_in_1d}
Here, we give an explicit construction of Green's function appearing in the TFD state for $D=1$ (QM). The four-point function (at large $N$) obeys the Schwinger--Dyson equation (see \cite{deMelloKoch:1996mj}) whose solution in one-dimensional momentum becomes
\begin{align}
\tilde{G}(n_1,n_2,n_3,n_4) = & \frac{2}{\beta^2}\delta_{n_1,n_3}\delta_{n_2,n_4}\phi(n_1)\phi(n_2) \nonumber \\
&-\frac{2c}{\beta^3}\delta_{n_1-n_2+n_3-n_4,0}\frac{\phi(n_1)\phi(n_2)\phi(n_3)\phi(n_4)}{1+\frac{c}{\beta}\sum_m\phi(m)\phi(n_2-n_1+m)}
\end{align}
with
\begin{equation}
\phi(n)=\frac{1}{\nu_n^2+\omega_\beta^2}.
\end{equation}
For the thermal four-point function we partition the time circle of inverse temperature $\beta$ into two halves obtaining the bi-time quantum fluctuation $\eta$ and its three components:
\begin{equation}
    \eta_{a,b} = \eta (\tau_a, \tau_b)
\end{equation}
with $\tau_1 = 0$, $ \tau_2 = \beta/2$.
Therefore, the equal-time thermal four-point function in coordinate space can be calculated from $\tilde{G}$ by Fourier transformation
\begin{equation}
\begin{split}
\langle\eta(0,0)\eta(0,0)\rangle &= \sum_{n_1,n_2,n_3,n_4}\tilde{G}(n_1,n_2,n_3,n_4)\\
&= \frac{1}{\omega x}\coth{\frac{\beta \omega}{2}}\coth{\frac{\beta x}{2}}+\frac{8\omega^3}{x^2}\frac{1}{4\omega^3(-1+\cosh{\beta\omega})+c(\beta\omega+\sinh{\beta\omega})},
\end{split}
\end{equation}
where we have defined
\begin{equation}
x=\sqrt{4\omega^2+\frac{c}{\omega}\coth{\frac{\beta\omega}{2}}}.
\end{equation}

As a consistency check, we serially expand the thermal four-point function in coupling constant $c$
\begin{align}
&\langle\eta(0,0)\eta(0,0)\rangle \nonumber \\
=&+ \frac{1}{2\omega^2}\coth^2{\frac{\beta\omega}{2}}-\frac{c}{128\omega^5}\operatorname{csch}^4{\frac{\beta\omega}{2}}(6\beta\omega+8\sinh{\beta\omega}+\sinh{2\beta\omega})+\frac{c^2}{8192\omega^8}\operatorname{csch}^4{\frac{\beta\omega}{2}}\times\nonumber \\
&
(-48+32\beta^2\omega^2+(-3+8\beta^2\omega^2)\cosh{\beta\omega}+48\cosh{2\beta\omega}+3\cosh{3\beta\omega}+108\beta\omega\sinh{\beta\omega})\nonumber \\
& + \mathcal{O}(c^3).
\end{align}
The zeroth order is the free propagator, which agrees with free ground state wavefunction. The first order is the interaction vertex, which agrees with coordinate space evaluation. The second order gives the one-loop correction.

Likewise, we can calculate the rest of thermal four-point functions. For $\langle\eta (0,0)\eta (0,\beta/2)\rangle$ we have
\begin{align}
&\langle\eta (0,0)\eta (0,\beta/2)\rangle = \sum_{n_1,n_2,n_3,n_4} (-1)^{-n_4} \tilde{G}(n_1,n_2,n_3,n_4) \nonumber \\
=&\frac{1}{\beta\omega}\operatorname{csch}{\frac{\beta\omega}{2}}\left(\frac{\beta}{2x}\coth{\frac{\beta x}{4}-\frac{2}{x^2}}\right)
+\frac{1}{4\beta\omega^3}\frac{2+\beta\omega\coth{\frac{\beta\omega}{2}}}{\sinh{\frac{\beta\omega}{2}}+\frac{c}{8\omega^3} \operatorname{csch}{\frac{\beta\omega}{2}}(\beta\omega+\sinh{\beta\omega})} .
\end{align}
For $\langle\eta (0, 0)\eta (\beta/2, \beta/2)\rangle$ we have
\begin{align}
&\langle\eta (0, 0)\eta (\beta/2, \beta/2)\rangle = \sum_{n_1,n_2,n_3,n_4} (-1)^{n_3-n_4} \tilde{G}(n_1,n_2,n_3,n_4) \nonumber \\
= & \frac{2}{\beta\omega}\coth{\frac{\beta\omega}{2}}\left(\frac{\beta}{2x}\operatorname{csch}{\frac{\beta x}{2}}-\frac{1}{x^2}\right)+\frac{1}{4\beta\omega^3}\frac{\beta\omega+\sinh{\beta\omega}}{\sinh^2{\frac{\beta\omega}{2}}+\frac{c}{8\omega^3}(\beta\omega+\sinh{\beta\omega})} .
\end{align}
For $\langle\eta (0, \beta/2)\eta (0, \beta/2)\rangle$ we have
\begin{align}
&\langle\eta (0, \beta/2)\eta (0, \beta/2)\rangle = \sum_{n_1,n_2,n_3,n_4} (-1)^{-n_2-n_4} \tilde{G}(n_1,n_2,n_3,n_4) \nonumber \\
=&
\frac{1}{2\omega^2}\coth^2{\frac{\beta\omega}{2}}+\frac{c}{4\omega^3 x}\operatorname{csch}^2{\frac{\beta\omega}{2}}\left(\frac{2\omega\coth{\frac{\beta x}{4}}-x\coth{\frac{\beta\omega}{2}}}{x^2-4\omega^2}+\frac{2}{\beta\omega x}\right) \nonumber \\
&
-\frac{c}{32\beta\omega^6}\frac{(2+\beta\omega\coth{\frac{\beta\omega}{2}})^2}{\sinh^2{\frac{\beta\omega}{2}}+\frac{c}{8\omega^3}(\beta\omega+\sinh{\beta\omega})} .
\end{align}
For $\langle\eta (0, \beta/2)\eta (\beta/2, 0)\rangle$ we have
\begin{align}
&\langle\eta (0, \beta/2)\eta (\beta/2, 0)\rangle = \sum_{n_1,n_2,n_3,n_4} (-1)^{-n_2+n_3} \tilde{G}(n_1,n_2,n_3,n_4) \nonumber \\
=&
\frac{1}{2\omega^2}\operatorname{csch}^2{\frac{\beta\omega}{2}}+\frac{c}{4\omega^3 x}\operatorname{csch}^2{\frac{\beta\omega}{2}}\left(\frac{2\omega\coth{\frac{\beta x}{4}}-x\coth{\frac{\beta\omega}{2}}}{x^2-4\omega^2}+\frac{2}{\beta\omega x}\right) \nonumber \\
&
-\frac{c}{32\beta\omega^6}\frac{(2+\beta\omega\coth{\frac{\beta\omega}{2}})^2}{\sinh^2{\frac{\beta\omega}{2}}+\frac{c}{8\omega^3}(\beta\omega+\sinh{\beta\omega})} .
\end{align}
These thermal four-point functions can further be assembled into a 3-by-3 $G$ matrix after symmetrization
\begin{align}
G_1 &= 2\langle\eta(0,0)\eta(0,0)\rangle \nonumber \\
G_2 &=\langle\eta (0, \beta/2)\eta (0, \beta/2)\rangle + \langle\eta (0, \beta/2)\eta (\beta/2, 0)\rangle \nonumber \\
G_3 &= 2\langle\eta (0,0)\eta (0,\beta/2)\rangle \nonumber \\
G_4 &= 2\langle\eta (0, 0)\eta (\beta/2, \beta/2)\rangle .
\end{align}
Putting the pieces together, we obtain the $G$ matrix for all coupling constants, $c$
\begin{equation}
G=
\begin{pmatrix}
G_{1} & G_{3} & G_{4} \\
G_{3} & G_{2} & G_{3} \\
G_{4} & G_{3} & G_{1} 
\end{pmatrix},
\end{equation}
Therefore, the interacting thermofield wavefunction at inverse temperature $\beta$ is given by
\begin{equation}
\Psi_\beta[\eta] =\frac{1}{\sqrt{Z}} \exp(-\frac{1}{2}\eta G^{-1}\eta) .
\end{equation}

One can verify that the $G$ matrix obeys the constraint equation in 3-by-3 matrix form
\begin{equation}
\label{eq:cs}
-G^{-1}KG^{-1} + V + \frac{c}{2}\Delta = 0 
\end{equation}
and satisfies the boundary condition
\begin{equation}
\lim_{c\rightarrow 0}G = G_0
\end{equation}
explicitly, where
\begin{equation}
K = \frac{1}{\omega_\beta}
\begin{pmatrix}
2\cosh{f} & \sinh{f} & 0 \\
\sinh{f} & 0 & -\sinh{f} \\
0 & -\sinh{f} & -2\cosh{f}
\end{pmatrix},
\end{equation}
\begin{equation}
V = 2 \omega_\beta^3
\begin{pmatrix}
\cosh{f} & -\sinh{f} & 0 \\
-\sinh{f} & 0 & \sinh{f} \\
0 & \sinh{f} & -\cosh{f}
\end{pmatrix},
\end{equation}
\begin{equation}
\Delta = 
\begin{pmatrix}
1 & 0 & 0 \\
0 & 0 & 0 \\
0 & 0 & -1
\end{pmatrix},
\end{equation}
and
\begin{equation}
G_0 = \frac{1}{2\omega_\beta^2}
\begin{pmatrix}
2\cosh^2{f} & \sinh{2f} & 2\sinh^2{f} \\
\sinh{2f} & \cosh{2f} & \sinh{2f} \\
2\sinh^2{f} & \sinh{2f} & 2\cosh^2{f} 
\end{pmatrix}.
\end{equation}
We used $\tanh{(f/2)} = \exp(-\beta\omega_\beta/2)$ in the expressions above.


\bibliographystyle{jhep}
\bibliography{reference}

\end{document}